\documentclass[10pt,twocolumn,prb,aps,floatfix,longbibliography]{revtex4-1}
\usepackage{color}
\usepackage{graphicx}
\usepackage[color=green!60]{todonotes}
\usepackage{physics}
\usepackage{amsthm}
\usepackage{amsmath}
\usepackage{amssymb}
\usepackage{enumerate}
\usepackage{placeins}
\usepackage{booktabs}
\usepackage{dsfont}
\usepackage{hyperref}

\newcommand{\ren}{R\'{e}nyi~}
\newcommand{\Eqref}[1]{Eq.~\eqref{#1}}
\AtBeginDocument{%
\heavyrulewidth=.08em
\lightrulewidth=.05em
\cmidrulewidth=.03em
\belowrulesep=.65ex
\belowbottomsep=0pt
\aboverulesep=.4ex
\abovetopsep=0pt
\cmidrulesep=\doublerulesep
\cmidrulekern=.5em
\defaultaddspace=.5em
}

\begin{document}

\title{Operationally accessible entanglement of one dimensional spinless fermions}

\author{Hatem Barghathi}
\affiliation{Department of Physics, University of Vermont, Burlington, VT 05405, USA}

\author{Emanuel Casiano-Diaz}
\affiliation{Department of Physics, University of Vermont, Burlington, VT 05405, USA}

\author{Adrian Del Maestro}
\affiliation{Department of Physics, University of Vermont, Burlington, VT 05405, USA}

\begin{abstract}
For indistinguishable itinerant particles subject to a superselection rule fixing their total number, a portion of the entanglement entropy under a spatial bipartition of the ground state is due to particle fluctuations between subsystems and thus is inaccessible as a resource for quantum information processing.  We quantify the remaining operationally accessible entanglement in a model of interacting spinless fermions on a one dimensional lattice via exact diagonalization and the density matrix renormalization group.  We find that the accessible entanglement exactly vanishes at the first order phase transition between a Tomonaga-Luttinger liquid and phase separated solid for attractive interactions and is maximal at the transition to the charge density wave for repulsive interactions.  Throughout the phase diagram, we discuss the connection between the accessible entanglement entropy and the variance of the probability distribution describing intra-subregion particle number fluctuations. 
\end{abstract}

\maketitle

\section{Introduction}

The entanglement of a quantum mechanical system can be exploited as a resource, allowing spatially separated parties to perform protocols (\emph{e.g.}~dense coding \cite{Bennett:1992dc}, teleportation \cite{Bennett:1993jc} and quantum cryptography \cite{Ekert:1991}) not feasible in a classical setting. The quantification of the exact amount of entanglement encoded in a given state is thus an important task that can be accomplished by studying the von Neumann entropy of a subsystem\cite{Schumacher:1995dg, Horodecki:2009gb}.  The situation can become more complicated in a condensed matter setting \cite{Laflorencie:2016hj}, especially when considering an eigenstate of some physical Hamiltonian governing a system of indistinguishable and itinerant interacting particles, whose total number is fixed.  Unlike an optical system of photons, conservation of total particle number $N$ for atoms or electrons may restrict the set of possible local operations, often referred to as a superselection rule (SSR) \cite{Wick:1952et}, and can potentially limit the amount of entanglement that can be physically accessed \cite{Horodecki:2000hr, Bartlett:2003ud, Schuch:2004jv, Kitaev:2004iy, Dunningham:2005fu,Marzolino:2015}.  This can be understood as originating from the fundamental inability to create coherent superposition states with different particle number in a subsystem.  As a result entanglement due to particle fluctuations alone cannot be utilized without access to a global phase reference \cite{Aharonov:1967be}.  In a pioneering work, Wiseman and Vaccaro \cite{Wiseman:2003jx} demonstrated that by averaging the von Neumann entanglement entropy of spatial modes over sectors corresponding to all possible numbers of particles in the subsystem defining those modes, they could place an upper bound on the amount of entanglement that could be transferred to a quantum register using local operations and classical communication.  This quantity, known as the \emph{accessible entanglement entropy}, has been previously studied for few-particle \cite{Wiseman:2003jx,Wiseman:2003vn,Vaccaro:2003ei, Dowling:2006kq, Herdman:2014ey, Melko:2016bo} or non-interacting \cite{Klich:2008se, Barghathi:2018oe} systems.  However, the interplay between interactions and an SSR fixing the total particle number has yet to be fully explored.  This is especially acute as many of the proposed or currently implemented quantum simulators \cite{Buluta:2009cq}, including those employing ultracold atoms \cite{Jaksch:2005go}, trapped ions \cite{Schmied:2008dh} and electrons \cite{Byrnes:2008mt, Mostame:2008sa}, are subject to fixed total $N$.

In this paper, we perform a systematic study of the accessible entanglement in an interacting model of spinless fermions, (the ``$t-V$ model''), on a one dimensional lattice, which is known to exhibit a host of interesting behavior \cite{DesCloizeaux:1966}, including first and second order quantum phase transitions between both classically ordered and quantum disordered phases.  We employ large scale exact diagonalization (ED) to study the ground state entanglement as a function of interaction strength for systems including up to $32$ sites at different filling fractions.  We compute both the originally defined von Neumann measure of accessible entanglement \cite{Wiseman:2003jx}, as well as its recently introduced \ren generalization \cite{Barghathi:2018oe}.  In order to investigate  the finite size scaling of the accessible entanglement near the quantum phase transition to the localized charge density wave state, we perform density matrix renormalization group (DMRG) calculations using the ITensor library \cite{itensor}.  

In the limits of infinitely strong repulsive and attractive interactions,  we derive analytical results for the accessible entanglement and find that in the thermodynamic limit, the accessible entanglement is constant and equal to $\ln 2$ at half-filling. At the interaction strength corresponding to the first order phase transition, the ground state is ``flat'' with all possible spatial occupations of the fermions contributing with equal weight.  Here, the accessible entanglement is identically zero at all filling fractions.  This result indicates that all of the entanglement between spatial subsystems at the transition is purely due to classical particle number fluctuations between subregions and thus the entanglement entropy is equivalent to the Shannon entropy of the corresponding probability distribution \cite{Klich:2008se}. This is a fermionic example of what was previously found in Bose-Einstein condensates and squeezed states of the Dicke model \cite{KlichRefaelSilva:2006}.  In the intervening quantum liquid,  where the microscopic system is described at low energies by Tomonaga-Luttinger liquid (TLL) theory, the accessible entanglement is reduced from the spatial entanglement for a spatial subregion of length $\ell$ by a subleading double log: $\sim\ln (K\ln \ell)$ where $K$ is the Luttinger parameter.  We confirm this asymptotic scaling \cite{Klich:2008se}, for finite sized systems by exploiting the exact solution of the $t-V$ model to obtain $K$ and determine that this behavior is predicated on the rapid convergence of the subsystem particle number probability distribution to a continuous Gaussian.  The discreteness of the local number of particles introduces corrections that are exponentially small in the width of the distribution which is substantial within the quantum liquid.  The accessible entanglement is maximal at the quantum phase transition between the TLL and charge density wave and appears to diverge in the thermodynamic limit signalling its potential use as a diagnostic measure more akin to a susceptibility than an order parameter.

The generalization of the operationally accessible entanglement to the \ren entropies described by an integer index $\alpha$ is of considerable interest, as these are amenable to measurement without access to the complete density matrix \cite{Calabrese:2004hl,Daley:2012bd}. Recent work identified the unique \ren generalization of accessible entanglement \cite{Barghathi:2018oe} and we have measured it via exact diagonalization for the ground state of the $t-V$ model.  We find that the reduction of entanglement due to the superselection rule fixing the total number of particles is well described by the classical \ren entropy of the subsystem particle number distribution.  This is not true in general, but approximately holds here due to a near proportionality between rescaled and bare number fluctuations. This proportionality is quantified and it is eventually violated for sufficiently large \ren indices. In the TLL phase where particle fluctuations between subregions are expected to be Gaussian, we explore the validity of a recent prediction for symmetry resolved entanglement \cite{Goldstein:2018kf} and find deviations that can be attributed to the amplification of finite size and ultraviolet cutoff effects for large \ren index $\alpha \sim 10$.

The main contributions of this work include: (1) confirmation that a system of fermions with fixed total particle number may act as a substantial entanglement resource for quantum information applications.  (2) The identification of putative power-law scaling of the exponential of the accessible entanglement entropy near the continuous quantum phase transition from a Tomonaga-Luttinger liquid to an insulator.  This transition thus identifies a critical coupling strength between fermions where the maximal amount of entanglement can be transferred to a quantum register. (3) By quantifying the role of the classical probability distribution governing the number of particles in a spatial subregion in placing a bound on the von Neumann and \ren generalized accessible entanglement entropies, we open up new experimental and computational avenues for the analysis of fermionic many-body phases as candidate resource states.

In the remainder of this paper, we provide a careful definition of the accessible entanglement entropy and discuss a few physical situations where its behavior is currently understood.  We then move on to the definition of the model in question, the $t-V$ model, and derive a number of new results in some analytically tractable limits.  The full phase diagram is then explored via ED and DMRG, where we answer the question of the exact amount of entanglement that can be extracted from a finite size system of interacting lattice fermions.  We identify the importance of the probability distribution controlling subsystem particle number occupation and conclude with a brief discussion on the effects of the finite system sizes under investigation and the role of the filling fraction.

\section{Accessible Entanglement}

\subsection{The \ren Entanglement Entropy}

The amount of entanglement that exists between some partition $A$ and its compliment $\bar{A}$ of a quantum many-body system in pure state $\ket{\Psi}$ can be quantified via the R\'{e}nyi entanglement entropy, which depends on an index $\alpha$ :
\begin{equation}
S_{\alpha} (\rho_A) = \frac{1}{1-\alpha}\ln \Tr\, \rho_A^{\alpha}
\label{eq:S_alpha}
\end{equation}
where $\rho_{A}$ is the reduced density matrix of partition $A$ obtained by
tracing out all degrees of freedom in $\bar{A}$ from the full density matrix:
\begin{equation}
    \rho_{A} = \Tr_{\bar{A}}\, \rho = \Tr_{\bar{A}} \ket{\Psi}\bra{\Psi}
\end{equation}
The \ren entropy is a non-increasing function of $\alpha$ and for $\alpha > 1$  
is bounded from above by the von Neumann entropy, $S_1(\rho_A) = -\Tr \rho_A \ln \rho_A$.

For a quantum many-body system subject to physical laws conserving some quantity (particle number, charge, spin, etc.), the set of local operations on the state $\ket{\Psi}$ are limited to those that don't violate the corresponding global superselection rule.  For the remainder of this paper, we will focus on our discussion on the case of fixed total $N$ and thus we are restricted to only those operators which locally preserve the particle number in $A$.  The effect this has on the amount of entanglement that can be transferred to a qubit register is apparent from the simple example (adapted from Ref.~\onlinecite{Wiseman:2003vn}) of one particle confined to two spatial modes $A$ and $\bar{A}$ corresponding to site occupations.  Then, for the state $\ket{\Psi} = \left(\ket{1}_A \otimes \ket{0}_{\bar{A}} + \ket{0}_A \otimes \ket{1}_{\bar{A}} \right)/\sqrt{2}$, Eq.~\eqref{eq:S_alpha} gives that $S_1 = \ln 2$. However, this entanglement cannot be transferred to a register prepared in initial state $\ket{0}_R$ via a $\texttt{SWAP}$ gate:
\begin{align*}
    & \texttt{SWAP} \ket{0}_R\otimes\left(\ket{1}_A \otimes
    \ket{0}_{\bar{A}} + \ket{0}_A \otimes \ket{1}_{\bar{A}} \right)/\sqrt{2} \\
    &= \frac{1}{\sqrt{2}}\left( \ket{1}_R \otimes \ket{0}_A \otimes
    \ket{0}_{\bar{A}} \ket{0}_R \otimes \ket{0}_A \otimes \ket{1}_{\bar{A}} \right)
\end{align*}
where the first term is not physically allowable due to the restriction that the number of particles in the system is fixed to be one. The post-swap result remains in a product state and the amount of transferable entanglement is identically zero.

\subsection{von Neumann Accessible Entanglement: $\alpha = 1$}

Thus, Eq.~\eqref{eq:S_alpha}, which includes the effects of non-local number fluctuations between $A$ and $\bar{A}$, overcounts the amount of entanglement that can be accessed from the system.  To quantify the physical reduction, Wiseman and Vaccaro \cite{Wiseman:2003jx} suggested that, for the case of $\alpha = 1$, a more appropriate measure should weight contributions to the entanglement coming from each superselection sector corresponding to the number of particles $n$ in $A$:
\begin{equation}
    S_1^{\rm{acc}}(\rho_A)=\sum_{n=0}^{N} P_n S_1(\rho_{A_n})
\label{eq:S1acc}.
\end{equation}
Here $\rho_{A_{n}}$ is defined to be the reduced density matrix of $A$, projected onto the subspace of fixed local particle number $n$ 
\begin{equation}
    \rho_{A_n} = \frac{1}{P_n}{\mathcal{P}}_{A_n} \rho_{A} {\mathcal{P}}_{A_n}
\label{eq:rhoAn}
\end{equation}
accomplished via a projection operator ${\mathcal{P}}_{A_n}$ that 
acts locally in partition $A$ fixing the number of particles in it to $n$ and the conservation of the total number of particles $N$ guarantees $N-n$ particles in its complement $\bar{A}$.  The probability of finding $n$ particles in $A$ is given by:
%
\begin{equation}
    P_n = \mathrm{Tr}\, {\mathcal{P}}_{A_n} \rho_A{\mathcal{P}}_{A_n}
    = \expval{{\mathcal{P}}_{A_n}}{\Psi}.
    \label{eq:Pn}
\end{equation}
As the projection constitutes a local operation which can only decrease
entanglement,  it is clear that $S_1^{\rm acc}(\rho_A) \le S_1(\rho_A)$. 
Moreover, the difference 
\begin{equation}
    \Delta S_1 (\rho_A) \equiv S_1(\rho_A) - S_1^{\rm acc}(\rho_A)
    \label{eq:DeltaS1}
\end{equation}
can be determined  by noting that the superselection rule guarantees that $[\rho_A,\hat{n}]=0$ where $\hat{n}$ is the number operator acting in partition $A$. Thus $\rho_A$ is block-diagonal in $n$ and it can be shown \cite{Klich:2008se} that 
\begin{equation}
    \Delta S_1 (\rho_A) =  H_1(\{P_n\})
\label{eq:DS1H1}
\end{equation}
where
\begin{equation}
    H_1(\{P_n\}) = -\sum_{n=0}^N P_n \ln P_n \le  \tfrac{1}{2}\ln\left(2\pi e \sigma^2 + \tfrac{1}{12}\right)
\label{eq:H1}
\end{equation}
is the Shannon entropy of the number probability distribution where 
\begin{equation}
    \sigma^2 \equiv \expval{n^2}-\expval{n}^2 = \sum_{n=0}^N n^2 P_n  - \left(\sum_{n=0}^N n P_n\right)^2\,.
\label{eq:sigma2}
\end{equation}
If $P_n$ is a discrete Gaussian distribution, $P_n \propto \exp[-(n-\expval{n}^2)/(2\sigma^2)]$ with $\expval{n} \gg \sigma \gg 1$, then the von Neumann entanglement entropy is reduced by an amount which only depends on the variance, $\Delta S_1 = \tfrac{1}{2} \ln(2\pi e \sigma^2)$.

\subsection{\ren Accessible Entanglement: $\alpha \ne 1$}
\label{ssec:RenyiAccEE}

Computing the accessible entanglement for a many-body system is a difficult task for $\alpha=1$, as full state tomography is required to reconstruct the density matrix $\rho$. However, for integer values with $\alpha > 1$ a replica trick can be used to recast $\mathrm{Tr} \rho_A^\alpha$ as the expectation value of some local operator \cite{Calabrese:2004hl}. This advance has led to a boon of new entanglement results using both computational \cite{Hastings:2010dc, Humeniuk:2012cq, McMinis:2013dp, Herdman:2014ey, Drut:2015fs} and experimental \cite{Daley:2012bd, Islam:2015cm, Kaufman:2016ep, Pichler:2016ec, Linke:2017tf, Lukin:2018wg} methods.  Motivated by this progress, two of us recently generalized the accessible entanglement to the case of \ren entropies with $\alpha \ne 1$ and found that \cite{Barghathi:2018oe}:
\begin{equation}
S_{\alpha}^{{\rm acc}} (\rho_A) = \frac{\alpha}{1-\alpha}\ln\left[ \sum_{n} P_n \mathrm{e}^{\frac{1-\alpha}{\alpha} S_{\alpha}(\rho_{A_n})}\right]
\label{eq:S_alpha_acc}
\end{equation} 
%
which reproduces Eq.~\eqref{eq:S1acc} in the limit $\alpha \to 1$. While not physically transparent in this form, the modification from the $\alpha=1$ case results from replacing the geometric mean in Eq.~\eqref{eq:S1acc} with a general power mean whose form is constrained by the physical requirement that
\begin{equation}
 0 \le \Delta S_\alpha \le \ln (N+1)
\label{eq:DeltaS_alpha_inq}
\end{equation}
where the upper bound is equal to the support of $P_n$. \Eqref{eq:S_alpha_acc} can also be interpreted as the quantum generalization of the conditional classical \ren entropy \cite{Cachin97entropymeasures,GolshaniPashaYari:2009,Hayashi:2011,SKORIC:2011el,FehrBerens2014}, subject to physical constraints \cite{Barghathi:2018oe}. The arguments leading to Eq.~\eqref{eq:DS1H1} can then be generalized (see the supplemental material of Ref.~[\onlinecite{Barghathi:2018oe}]) leading to 
\begin{equation}
    \Delta S_{\alpha}\equiv  S_{\alpha} - S_{\alpha}^{{\rm acc}} = H_{{1}/{\alpha}}\left(\{P_{n,\alpha}\}\right)
\label{eq:S_alpha_acc5}
\end{equation}
where we introduce the classical \ren entropy of $P_n$
\begin{equation}
    H_{\alpha}\left(\{P_n\}\right)=\frac{1}{1-\alpha}\ln\sum_n P_n^{\alpha} 
\label{eq:Halpha}
\end{equation}
and
\begin{equation}
    P_{n,\alpha}=\frac{\Tr\, \left[{\mathcal{P}}_{A_{n}} \rho_A^{\alpha} 
    {\mathcal{P}}_{A_{n}}\right]}{\Tr\, \rho_A^{\alpha}}=\frac{P_n^\alpha\Tr\, \rho_{A_n}^{\alpha}}{\Tr\, \rho_A^{\alpha}}
\label{eq:Pna}
\end{equation}
can be interpreted as a normalization of partial traces of $\rho_A^{\alpha}$, where the SSR fixing the total particle number leads to $\Tr\, \rho_A^{\alpha}=\sum_n \Tr\, \left[{\mathcal{P}}_{A_{n}} \rho_A^{\alpha} {\mathcal{P}}_{A_{n}}\right]$ and thus guarantees the normalization of $P_{n,\alpha}$. Note that we have defined $P_{n,1} \equiv P_n$ for notational consistency. For brevity, let $H_{\alpha}(\{P_n\}) \equiv H_{\alpha}$ from here onwards. 

Writing the difference $\Delta S_{\alpha}$ as the classical \ren entropy of the fictitious probability distribution $P_{n,\alpha}$, simplifies the calculation of $\Delta S_{\alpha}$ and clarifies its properties, e.g., the fact that $H_{\alpha}$  is positive and bounded from above by $ H_{0}=\ln(N+1)$ guarantees that $\Delta S_{\alpha}$ satisfies the physical requirement in \Eqref{eq:DeltaS_alpha_inq}. \cite{Barghathi:2018oe} In addition, $P_{n,\alpha}$ is fully determined by $P_n$ and the full and the projected traces of $\rho_A^{\alpha}$, i.e.~$\Tr\, \rho_{A}^{\alpha}$ and $\Tr\, \rho_{A_n}^{\alpha}$, which can be measured using the experimental and numerical methods mentioned above.    

Before proceeding to a discussion of previous results for the accessible entanglement entropy, let us consider the special case where the probability distribution $P_{n,\alpha} \propto (P_n)^\alpha$. Then, using Eq.~\eqref{eq:S_alpha_acc5} we have:
\begin{align}
    \Delta S_\alpha 
                    &= \frac{1}{1-\alpha^{-1}} \ln \sum_n \qty(\frac{P_n^\alpha}{\sum_n P_n^\alpha})^{1/\alpha} \nonumber \\
                    &= H_\alpha \ ,  
\label{eq:PnaPaHa}
\end{align}
which reproduces the von Neumann result in Eq.~\eqref{eq:DS1H1}.

\subsection{Previous Results}
While the accessible entanglement entropy can be used to diagnose the feasibility of using a many-body state of quantum matter as an entanglement resource, exact results  are mostly limited to non-interacting systems.  For a condensate of free bosons, the projected reduced density matrix $\rho_{A_n}$ is always pure for any $n$ and thus the accessible entanglement is zero \cite{Herdman:2014ey}.  For free fermions, early calculations \cite{Dowling:2006kq} found $S_1^{\rm acc} \ne 0$ in a thermal state under a non-contiguous spatial bipartition of two sites on a one-dimensional lattice.  More recent work on non-interacting spinless fermions \cite{Klich:2008se,Barghathi:2018oe} found that the SSR fixing the total particle number reduces the accessible entanglement by an amount that is subleading in the size of the spatial bipartition $\ell$ when $\ell \gg 1$. This result hinges on the realization that the probability distribution $P_{n,\alpha}$ defined in \Eqref{eq:Pna} is Gaussian with an average that is independent of $\alpha$ and a variance $\sigma_\alpha^2$ that scales as $\ell^{d-1}\ln\ell/\alpha$ in $d$ spatial dimensions.   As the spatial entanglement $S_\alpha$ scales as $\ell^{d-1}\ln\ell$, \cite{Leschke:2014el} $\Delta S_{\alpha}/ S_{\alpha}\sim \ln\left(\ell^{d-1}\ln\ell\right)/\left(\ell^{d-1}\ln\ell\right)$ which vanishes as $\ell \to \infty$.

For critical systems in $1d$ described by Luttinger liquid theory, (or more generally any conformal field theory with a conserved U(1) current), the particle number probability distribution $P_n$ is also asymptotically Gaussian with a variance $\sigma^2=K\ln\ell/\pi^2$ in the limit $\ell\gg1$ \cite{Klich:2008se, Song:2010eq, Song:2012cp,Calabrese:2012fa}, where $K$ is the Luttinger parameter. Here, a result by Goldstein and Sela \cite{Goldstein:2018kf} can be employed to investigate $P_{n,\alpha}$, which is Gaussian, having the same average as $P_n$ but with modified variance: $\sigma^2_\alpha= {\sigma^2}/{\alpha} \stackrel{\ell \gg 1}{\to} {K}/{\alpha\pi^2} \ln\ell$.  As a result
%
\begin{align}
\left. \Delta S_\alpha  \right \rvert_{\rm TLL} &= H_{{1}/{\alpha}}\left(\{ P_{n,\alpha}\}\right)=H_\alpha \nonumber \\
                    &=\frac{1}{2}\ln\sigma^2 +\frac{1}{2}\ln\qty[ 2\pi\alpha^{1/\left(\alpha-1\right)}]
\label{eq:gaussianEntropy}.
\end{align}
\Eqref{eq:gaussianEntropy} can be combined with the known result for the spatial entanglement entropy of a critical $1d$ system \cite{Holzhey:1994kc, Calabrese:2004hl, Calabrese:2009co}
\begin{equation}
    \left.  S_\alpha\right \rvert_{1d\,\rm CFT}=\frac{c}{6}\qty(1 + \frac{1}{\alpha})\ln\frac{\ell}{a_0} + {O}(1)
\label{eq:SalphaCFT},
\end{equation}
where $c$ is the central charge and $a_0$ is a short distance cutoff, to see that the 
fraction of non-accessible entanglement entropy $\Delta S_{\alpha}/ S_{\alpha}$, vanishes asymptotically as $\ln(\ln\ell)/\ln\ell$.

Studies of the interaction dependence of $S^{\rm acc}$ have been previously limited to bosonic systems in $1d$.  Quantum Monte Carlo simulations of 
harmonically trapped and harmonically interacting bosons identified a maxima in
the accessible entanglement as a function of interaction strength
\cite{Herdman:2014ey}. Exact diagonalization of the $1d$ Bose-Hubbard model at
unit filling for systems of up to $N=16$  demonstrated that $S_2^{\rm acc}$
vanishes in the limit of strong and weak interactions.\cite{Melko:2016bo}
Interestingly, $S_2^{\rm acc}$ was maximal near the superfluid-insulator phase
transition and appeared to obey phenomenological scaling for the limited system
sizes that could be studied. For an extended Bose-Hubbard model of four modes
that includes pair-correlated hopping,  exact diagonalizaition and variational
calculations identified an interesting regime with strong pair-correlations
where a matter wave beam-splitter operation on the ground state results in all
entanglement being accessible \cite{Volkoff:2019ga}.

Missing from this list is any system of interacting fermions and we now present
numerical results for spinless fermions in one spatial dimension.

\section{The $t-V$ Model of Interacting Spinless Fermions}
\label{sec:III}
%
\begin{figure*}[thp]
\begin{center}
\includegraphics[width=1.0\textwidth]{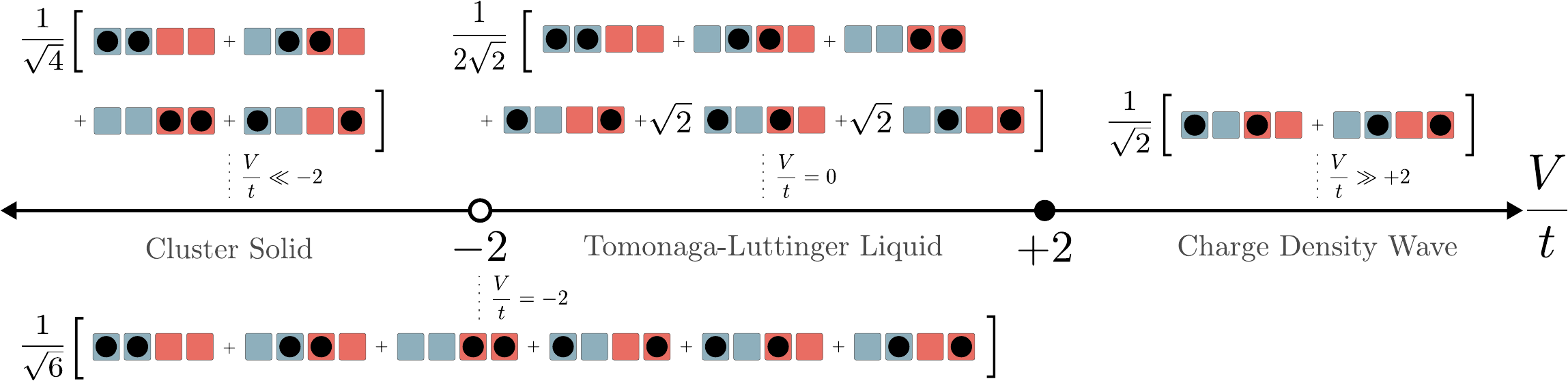}
\end{center}
\caption{Phase diagram of the $t-V$ model accompanied by pictures of candidate ground states for $N=2$ fermions on a $L=4$ site lattice with anti-periodic boundary conditions. For the purposes of measuring accessible entanglement, the lattice has been bipartitioned into spatial subregions $A$ (blue) and $\bar{A}$ (red), each of size $\ell = 2$. In the limit of strong attractive interactions where $V/t \ll -2$, the particles cluster together and there are $L$ equally probable configurations corresponding to all translations of the cluster.  At the first order phase transition where $V/t = -2$, all ${L}\choose{N}$ configurations are equally probable resulting in a flat state. In the TLL phase with $|V/t| < 2$,  particles are delocalized and we have included a characteristic state corresponding to free fermions $(V=0)$. In the limit of strong repulsive interactions where $V/t \gg 2$, fermions maximize their distance from each other resulting in a charge density wave (CDW) phase. The open and closed circles on the $V/t$ axis denote a first order and continuous phase transition, respectively.}
\label{fig:phaseDiagram}
\end{figure*}

\subsection{Description and Solution}

To investigate the behavior of accessible entangelment in an interacting fermionic system, we consider the $t-V$ model defined by a one-dimensional lattice of $L$ sites occupied by $N$ spinless fermions and governed by the Hamiltonian
\begin{equation}
    H= -t\sum_{i}\left(c^\dagger_{i} c^{\phantom{\dagger}}_{i+1} + \text{h.c.}\right) +V\sum_i n_i n_{i+1}\,
\label{eq:H-tV}
\end{equation}
where $c^\dagger_{i}$ and $ c^{\phantom{\dagger}}_{i}$ denote the fermionic creation and annihilation operators at site $i$, $\{c^{\phantom{\dagger}}_i,c_j^\dagger\}=\delta_{i,j}$, and $n_i=c^\dagger_{i}c^{\phantom{\dagger}}_{i}$.  Here, $t>0$ and $V$ represent the nearest-neighbor hopping amplitude and interaction strength, respectively. We consider a half-filled lattice ($L=2N$), unless mentioned otherwise and we use periodic boundary conditions (PBC) for an odd $N$, while for even $N$ we use antiperiodic boundary conditions (APBC) to avoid complications arising from the degenerate ground state.

Eq.~(\ref{eq:H-tV}) can be mapped onto the XXZ spin-1/2 chain (at fixed magnetization) which is exactly solvable via Bethe ansatz \cite{Lieb:1961dn, DesCloizeaux:1966, DesCloizeauxGaudin:1966} (see e.g.~[\onlinecite{Franchini:2017ky}] for a recent pedagogical review). For  $-2<V/t< 2$, at low energies and long wavelengths, the system can be understood as a Tomonaga-Luttinger liquid where the TLL parameter $K$ at half filling, is \cite{Haldane:1980ew} 
\begin{equation}
 K= \frac{\pi}{2\cos^{-1}\left[-V/(2t)\right]}\,.
\label{eq:KtV}
\end{equation}
In this language, $0 < K < 1$ corresponds to repulsive $(V>0)$ interactions, $K>1$ to attractive $(V<0)$ interactions, and non-interacting fermions $(V=0)$ have $K = 1$. By increasing the relative interaction strength $|V/t|$, the system undergoes two phase transitions, a first order phase transition to a single fermionic cluster phase at $V/t=-2$, $(K = \infty)$ and a continuous one at $V/t=2$, $(K=1/2)$ to charge-density wave (CDW) phase. A schematic phase diagram is shown in Fig.~\ref{fig:phaseDiagram}.

\subsection{Exact Ground State Results For Accessible Entanglement}

In this section we derive a number of exact and asymptotic results for the accessible entanglement entropy of the $t-V$ model using insights gained from the structure of the ground states depicted in Fig.~\ref{fig:phaseDiagram}. Results for the von Neumann accessible entanglement are summarized in Table~\ref{tab:Limits}. 
\begin{table}[htp]
\begin{center}
   \renewcommand{\arraystretch}{1.8}
   \setlength\tabcolsep{8pt}
 \begin{tabular}{@{}lll@{}} 
  \toprule
    Interaction	& $S_{1}^{{\rm acc}} (\rho_A)$	& $\Delta S_{1}$ \\ 
   \midrule
    ${V}/{t} \to  \infty$ &  ${\tfrac{1}{2}\qty[1+(-1)^\ell]\ln 2}^\dag$ &  ${\tfrac{1}{2}\qty[1-(-1)^\ell]\ln 2}^\dag$	\\  
    ${V}/{t} \to -\infty$ &  $\tfrac{L-2}{L}\ln 2$ & $\ln\tfrac{L}{2}^\natural$ \\ 
     $V/t =-2$ &  $0^{\dag \ddag}$ & $\frac{1}{2}\ln L^\natural$ \\
    \bottomrule
    \end{tabular}
\end{center}
\caption{\label{tab:Limits} 
Analytical results for the accessible entanglement in the ground state of the $t-V$ model with $N$ fermions on $L$ sites under a spatial bipartition consisting of $\ell=L/2$ contiguous sites. Symbols indicate approximations or generalizations with $\natural$ marking that the expression is asymptotically valid in the limit $L \gg 1$, $\dag$ means $\ell<L$ and $\ddag$ that the result is true for any filling fraction $N<L$.}
\end{table}

\subsubsection{$V/t \to \infty$}

In the limit $V/t \to \infty$ and at half filling, the system reduces its energy by separating every two fermions by at least one empty site and thus the ground state $|\Psi_{V/t \to \infty}\rangle = \left(\ket{\psi_{\rm{even}}} + \ket{\psi_{\rm odd}}\right)/\sqrt{2}$ is an equal superposition of two occupation states. In one state the fermions occupy sites with only even indices ($\ket{\psi_{\rm{even}}}=\ket{0101 \dots 0101}$) and in the other, they occupy sites with only odd indices $\ket{\psi_{\rm{odd}}}=\ket{1010 \dots 1010}$).

If we now consider spatial bipartition $A$ consisting of $\ell$ consecutive sites, we can write 
\begin{align}
    \ket{\Psi_{V/t \to \infty}} &= \frac{1}{\sqrt{2}}  \ket{\psi_{\rm{even}}}_A\otimes\ket{\psi_{\rm{even}}}_{\bar{A}} \nonumber \\
\qquad\quad &+ \frac{1}{\sqrt{2}}\ket{\psi_{\rm{odd}}}_A\otimes\ket{\psi_{\rm{odd}}}_{\bar{A}}, 
\end{align}
resulting in the reduced density matrix 
\begin{equation*}
    \rho_A = \frac{1}{2}\ket{\psi_{\rm{even}}}_A\bra{\psi_{\rm{even}}}+\frac{1}{2}\ket{\psi_{\rm{odd}}}_A\bra{\psi_{\rm{odd}}}. 
\end{equation*}
For even $\ell$, both of $\ket{\psi_{\rm{even}}}_A$ and $\ket{\psi_{\rm{odd}}}_A$ represent $\ell/2$ fermions as the number of sites with odd indices is equal to the number of sites with even indices. Therefore, the number of particles in partition $A$ is fixed to $\ell/2$ and the entanglement entropy of the projected state $\mathcal{P}_{A_{n=\ell/2}}\ket{\Psi_{V/t \to \infty}}$ is $S_{\alpha}(\rho_{A_{n=\ell/2}})=\ln2$ with $P_{n=\ell/2}=1$ resulting in an overall accessible entanglement entropy $S^{\rm{acc}}_{\alpha}(\rho_{A})=\ln2$. The picture is different for odd $\ell$ where the number of sites with odd indices differs from the number of sites with even indices by $1$. In this case one of the states $\ket{\psi_{\rm{even}}}_A$ and $\ket{\psi_{\rm{odd}}}_A$ will represent  $(\ell-1)/2$ fermions while the other represents $(\ell+1)/2$ fermions and therefore the projected state $\mathcal{P}_{A_{n=(\ell\pm1)/2}}\ket{\Psi_{V/t \to \infty}}$ is a separable state yielding zero entanglement entropy $S^{\rm{acc}}_{\alpha}(\rho_{A})=0$. For any partition size $\ell$, regardless of its parity, the spectrum of $\rho_{A}$ consists of two equal eigenvalues fixing the spatial entanglement entropy $S_{\alpha}(\rho_{A})$ to $\ln 2$.

\subsubsection{$V/t \to -\infty$}

In the other extreme, $V/t \to -\infty$ and for any number of fermions $0<N<L$, the system minimizes its energy by forming a cluster of fermions that extends over any consecutive $N$ sites. The ground state of the system, in this case, is an equal superposition of all $L$ possible clusters. Once more, considering a partition $A$ of $\ell$ consecutive sites, we can write the ground state as 
\begin{equation}
    \ket{\Psi_{V/t \to -\infty}} = \frac{1}{\sqrt{L}}\sum_n\sum_{i,j} \ket{n,i}_A\otimes\ket{N-n,j}_{\bar{A}}\, , 
\end{equation}
where $\ket{n,i}_A$ is the $i^{\rm{th}}$ configuration having $n$ particles in partition $A$ and $\ket{N-n,j}_{\bar{A}}$ is the $j^{\rm{th}}$ configuration with $N-n$ particles in its spatial compliment $\bar{A}$. Since the state is a superposition of $L$ particle configuration states, $\rho_{A}$ can have at most $L$ non-zero eigenvalues. This defines an upper bound on $S_{\alpha}(\rho_A) \le \ln L$.

The simplicity of the state  $\ket{\Psi_{V/t \to -\infty}}$ allows us to classify the projected state $\mathcal{P}_{A_n}\ket{\Psi_{V /t\to -\infty}}$ that corresponds to having $n$ particles in partition $A $ as follows. If the state $\mathcal{P}_{A_n}\ket{\Psi_{V/t \to -\infty}}$ has partition $A$ or its complement $\bar{A}$ either empty or fully occupied then $\ket{\Psi_{V/t \to -\infty}}$ must be a separable state with $S_{\alpha}(\rho_{A_{n}})= 0$.  What remains are the projected states $\mathcal{P}_{A_n}\ket{\Psi_{V/t \to -\infty}}$ in which both of $A$ and $\bar{A}$ have at least one empty and one occupied site. Due to the existence of the fermion cluster,  knowing the configuration of the $n$ particles in partition $A$ fully determines the configuration of the $N-n$ particles in partition $\bar{A}$. Moreover, there can be only two such configurations that correspond to the fermionic cluster emerging into the partition $A$ --  either from its left or right end, such that $\mathcal{P}_{A_n}\ket{\Psi_{V/t \to -\infty}}=\frac{1}{\sqrt{L}}\sum_{i=1}^2 \ket{n,i}_A\otimes\ket{N-n,i}_{\bar{A}}$, where $\braket{n,1}{n,2}_A=\braket{N-n,1}{N-n,2}_{\bar{A}}=0$. This gives $S_{\alpha}(\rho_{A_{n}})= \ln 2$ and $P_n=2/L$. A simple counting then gives the number of projected states $m$ that yield non zero entanglements as $\min\{\ell,L-\ell,N,L-N\}-1$. The resulting accessible entanglement is given by
\begin{equation}
S_{\alpha}^{{\rm acc}} (\rho_A) = \dfrac{\alpha}{1-\alpha}\ln\left[ \dfrac{2m}{L}2^{\frac{1-\alpha}{\alpha}}+1-\dfrac{2m}{L}\right]
\label{eq:S_alpha_neg_Inf},
\end{equation}
which simplifies to 
\begin{equation}
S_1^{\rm{acc}}(\rho_A)=\frac{2m}{L}\ln2
\label{eq:S_1_neg_Inf},
\end{equation}
in the von Neumann case $\alpha = 1$.  From \Eqref{eq:S_alpha_neg_Inf} we see that $S_{\alpha}^{{\rm acc}} (\rho_A)$ is an increasing function of $m$. For a given $L$, the maximum value of $m$ is $L/2-1$ which is achieved for $\ell=N=L/2$. In this case $S_1^{\rm{acc}}(\rho_A)=\frac{L-2}{L}\ln2$  and for $L\gg1$ we can write $S_{\alpha}^{{\rm acc}} (\rho_A) \approx\ln2$.

To calculate the spatial entanglement entropy of this state, in general, we need the full spectrum of $\rho_{A}$. Based on the above there will be $2m$ eigenvalues of $\rho_{A}$ that are equal to $1/L$. In addition, there are two more eigenvalues which correspond to one of the partitions being either empty or fully occupied. Counting the number of such occupation states gives the eigenvalues $\left(|\ell-N|+1\right)/L$ and $\left(|\ell+N-L|+1\right)/L$. Now if we consider the conditions for maximizing $S_{\alpha}^{{\rm acc}} (\rho_A) $, i.e., at half-filling and with half-partition, we find that $\rho_{A}$ has a flat spectrum with $L$ eigenvalues and thus $S_{\alpha}(\rho_A) $ is saturated at its upper bound, $S_{\alpha}(\rho_A) = \ln L$,  and therefore $\Delta S_{\alpha}\approx\ln\left(L/2\right)$, for $L \gg 1$.

\subsubsection{$V/t = -2$}
\label{FPT}

Now we turn our attention to the very interesting case of the first order phase transition at $V/t=-2$, where the ground state $\ket{\Psi_{V/t =-2}}$ is an equal superposition of all ${L}\choose{N}$ possible configurations of $N$ fermions on $L$ sites. (see Appendix~\ref{Appendix:A} for proof). In the language of the XXZ model this corresponds to the isotropic ferromagnetic point \cite{Faddeev1984}.  If we project $\ket{\Psi_{V/t=-2}}$ into a state with $n$ particles in partition $A$ and in one of its ${\ell}\choose{n}$ possible configurations, we get an equal superposition of ${L-\ell}\choose{N-n}$ occupation states which differ only by the configuration of the $N-n$ particles in $\bar{A}$. Therefore, we can immediately construct the desired Schmidt decomposition by inspection:
\begin{equation}
|\Psi_{V/t =-2}\rangle=\dfrac{1}{\sqrt{{L}\choose{N}}}\sum_{n}\sqrt{{{\ell}\choose{n}}{{L-\ell}\choose{N-n}}} |n\rangle_{A}\otimes|N-n\rangle_{\bar{A}}
\label{eq:Psi_neg_2}.
\end{equation}
Here, each of the normalized states $|n\rangle_{A}$ and $|N-n\rangle_{\bar{A}}$ is an equal superposition of all of the possible configurations of $n$ and $N-n$ particles in partitions $A$ and $\bar{A}$, respectively. 

For the state above, the projected reduced density matrix $\rho_{A_n}=|n\rangle_{A}\langle n|_{A}$ is a pure state and thus, for any $n$, $S_{\alpha}(\rho_{A_n})=0$. As a result, for any partition size $\ell$, the  accessible entanglement $S_{\alpha}^{{\rm acc}} (\rho_A)=0$.
Moreover, the spectrum of $\rho_{A}$ is given by the particle number probability distribution $P_n={{\ell}\choose{n}}{{L-\ell}\choose{N-n}}/{{L}\choose{N}}$ where we have used the fact that the block-diagonal structure of $\rho_{A}$ in $n$ allows us to write $\rho_{A}=\sum_n P_n \rho_{A_n}$. Furthermore, for this state, $S_{\alpha}(\rho_{A})=H_{1/\alpha}(\{ P_{n,\alpha}\}) =H_{\alpha}(\{P_n\})$

Let us consider the behavior of $S_{\alpha}(\rho_{A})$ in the limit $L \gg 1$ and, for clarity, we focus on an equal bipartition at half-filling: $\ell=N=L/2$. Here, $P_n={{\ell}\choose{n}}^2/{{2\ell}\choose{\ell}}$ and asymptotically it is a Gaussian distribution in $n$ with variance $\sigma^2=L/16$ and thus $S_{\alpha}(\rho_{A})=\Delta S_{\alpha}=\frac{1}{2}\ln\left( 2\pi\sigma^2\alpha^{\left(\alpha-1\right)^{-1}}\right)\approx\frac{1}{2}\ln{L}$.

\section{Numerical Results}

To test the validity of the predictions in the previous section, we calculate the accessible entanglement in the ground state of the  $t-V$ model,  defined in section~\ref{sec:III}, via numerical exact diagonalization for small systems (up to $32$ sites) and using DMRG for larger systems (up to $98$ sites),  where the calculations are performed using the ITensor C++ library \cite{itensor}. We focus on half-filling ($N=L/2$) and with a spatial partition size of $\ell=L/2$ contiguous sites, unless otherwise noted.  All data, code and scripts used in this paper can be found online \cite{repo}. 

\begin{figure}[t]
\begin{center}
\includegraphics[width=1.0\columnwidth]{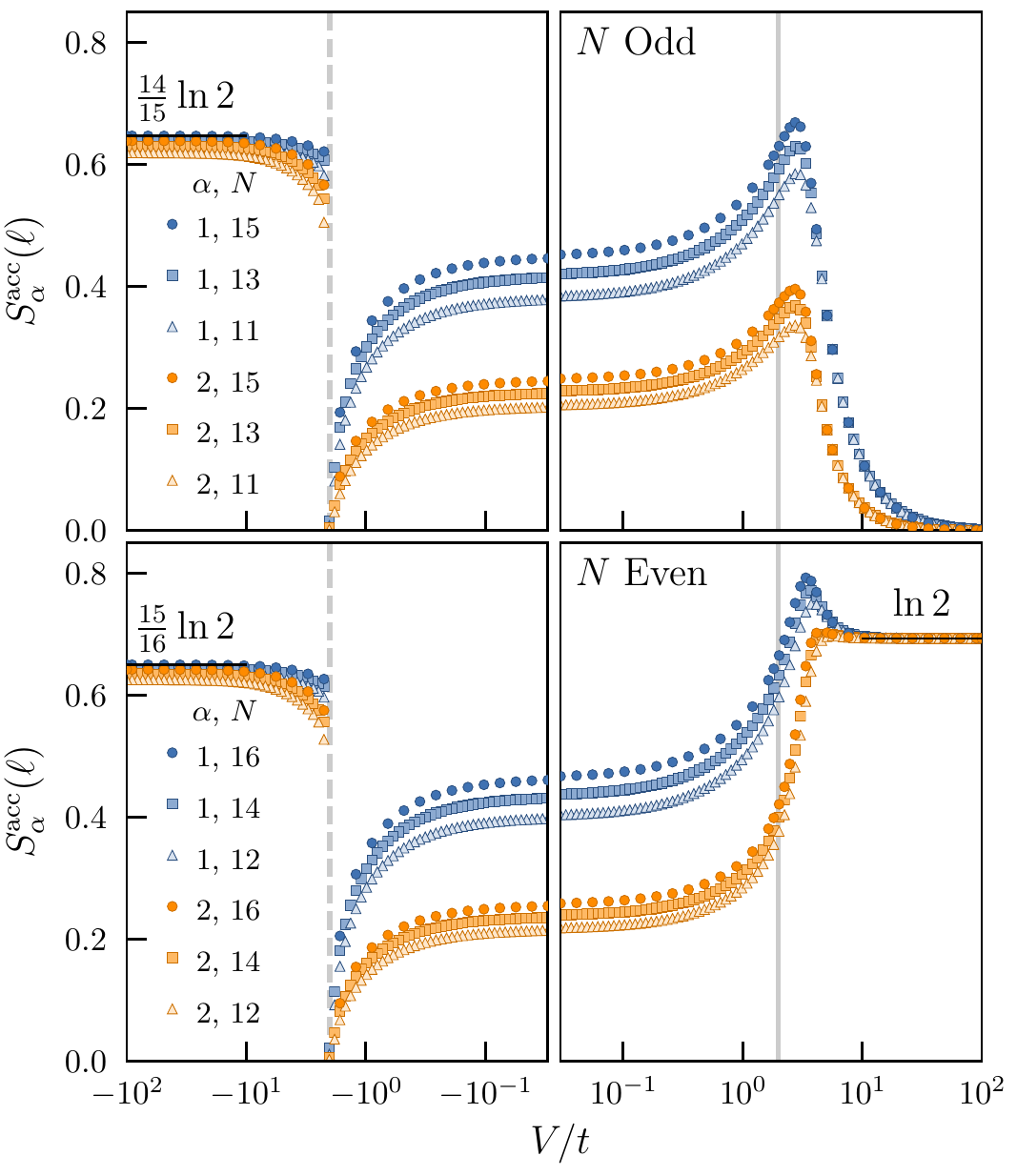}
\end{center}
\caption{Accessible entanglement entropy $S_{\alpha}^{\mathrm{acc}}(\ell)$ for $\alpha = 1, 2$ in the ground state of the $t-V$ model as a function of interaction strength $V/t$ at half-filling, $N = L/2$. The top panel shows the results for an odd number of total particles: $N=11,13,15$ and the bottom, for even: $N=12,14,16$. The solid and dashed gray vertical lines indicate the locations of the known phase transitions for the model, $V/t = \pm 2$. For $N=15,16$ the asymptotic results computed in Section~\ref{sec:III} in the limits $V/t \to \pm \infty$ for $S_{1}^{\mathrm{acc}}$ are shown as solid black lines.}
\label{fig:OEE}
 \end{figure}

Figure~\ref{fig:OEE} shows the von Neumann and second \ren accessible entanglement entropies, $S_{1}^{\mathrm{acc}}$ and $S_{2}^{\mathrm{acc}}$, as a function of the dimensionless interaction strength $-100 \le V/t \le 100$ for the six largest system studied by ED\@. To illustrate the effects that the parity of $N$ has on $S_{\alpha}^{\mathrm{acc}}$, the top and bottom panels of Fig.~\ref{fig:OEE} correspond to odd and even $N$, respectively.
%
%

\subsection{Phase transitions and limiting cases of $V/t$}

Starting from the regime of strong attractive interactions, $V/t =-100$, in Fig.~\ref{fig:OEE},  we see that $S_{1}^{{\rm acc}} (\rho_A)$ is rapidly converging to the expected value $(1-1/N)\ln 2$ in the limit $V/t \to -\infty$ (Eqs.~\eqref{eq:S_alpha_neg_Inf} and \eqref{eq:S_1_neg_Inf}).  This asymptotic result persists down to nearly $V/t = -10$ for large system sizes.  Increasing $V/t$ further, $S_{\alpha}^{{\rm acc}} (\rho_A)$ decreases slowly until we get closer to the first order phase transition at $V/t=-2$, (see section~\ref{FPT}.), where $S_{\alpha}^{{\rm acc}} (\rho_A) $ decreases rapidly until it vanishes exactly at the transition point.  This result holds for all $N$.  As we increase $V/t $ beyond $-2$,  $S_{\alpha}^{{\rm acc}} (\rho_A) $ grows in the TLL regime as interaction driven liquid correlations build up until it eventually peaks in the vicinity of the infinite system critical point ($V/t=2$) and eventually saturates to its limiting $V/t \to\infty$ value by $V/t \simeq 100$ which depends on the particle number parity: $S_{\alpha}^{\rm acc} \to 0$ for $N$ odd and $S_{\alpha}^{\rm acc} \to \ln 2$ for $N$ even. Exact diagonalization results up to $L = 32$ sites indicate that finite size effects are most visible in the Tomonaga-Luttinger liquid phase and this is especially true as we approach the continuous phase transition at $V/t = 2$ where a maxima begins to develop in the accessible entanglement entropy.

Quantum information measures have been known for some time to show signatures
at continuous and discrete phase transitions, both at $T=0$ and finite
temperature \cite{Osterloh:2002oq, Osborne:2002tu, Gu:2003jg, Verstraete:2004rq,Somma:2004ie, Anfossi:2005il,Larsson:2005qt, Popp:2005ex, Iaconis:2013vg, Iemini:2015dh, Yuste:2018fo, Lu:2019hs, Walsh:2019mc,Braun:2018} including the case of spinless fermions under consideration here \cite{Gu:2003jg, Ren:2012ms, Zheng:2015lt, Chen:2006sv, Zhang:2018kv}. A commonality amongst these studies is that the information quantity in question (entanglement entropy, negativity, concurrence, purity, etc.) develops some feature akin to an order parameter.  Here, an analysis of the exact diagonalization data shows that the accessible entanglement develops a maximum at a coupling strength $\tfrac{V}{t}\rvert_{\rm max} \sim \mathrm{O}(1)$. Making the empirical observation that the accessible entanglement appears to behave like a susceptibility, we perform an analysis of how the distance of the maxima from the infinite system size critical point ($\delta = \tfrac{V}{t}\rvert_{\rm max} - 2$) depends on the system size $L$ to search for power law scaling.

%
\begin{figure}[t]
\begin{center}
\includegraphics[width=1.0\columnwidth]{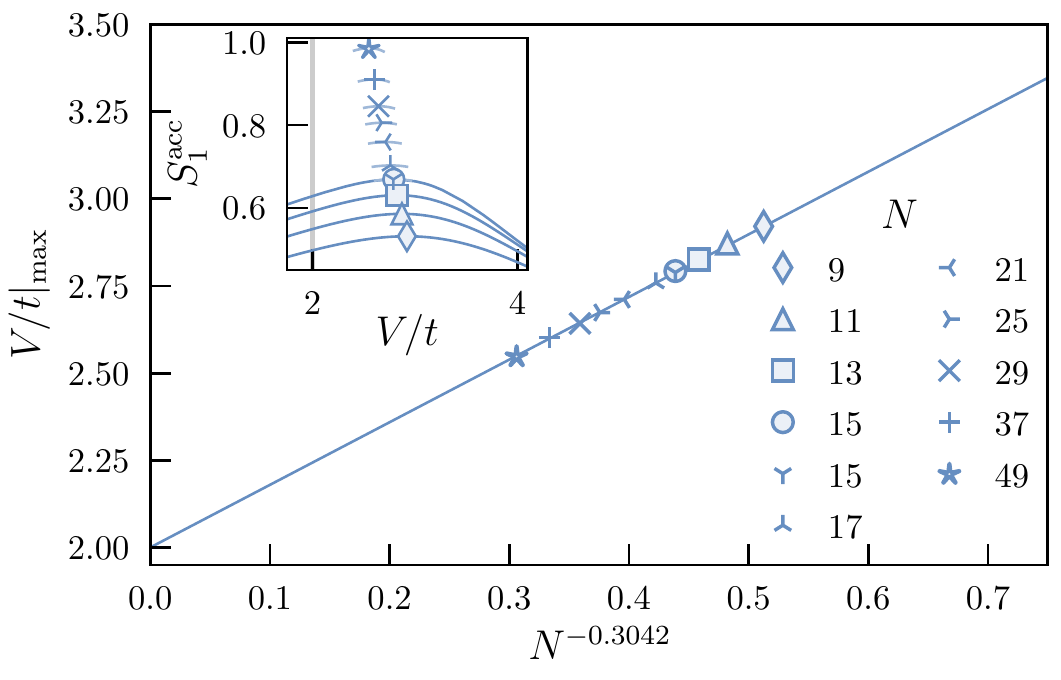}
\end{center}
\caption{Interaction strength at which the maximum $S_{1}^{\mathrm{acc}}$ occurs as a function of the total number of particles $N$. Filed shapes $(N\le 15)$ were computed with exact diagonalization while the remaining symbols for $N\ge 15$ are the result of density matrix renormalization group calculations.  Finite size corrections are investigated via a 2-parameter fit of $\ln N$ vs. $\ln{(V/t - 2)}$ and support scaling toward the infinite size phase transition at $V/t = 2$.  Inset: The interaction dependence of $S_{1}^{\mathrm{acc}}$ for various $N$ in the neighborhood of $\tfrac{V}{t}\rvert_{\rm max}$ shows an evolving peak.}
\label{fig:peakScalingOddN}
\end{figure}
%
In order to investigate the existence of such scaling using larger system sizes
than are possible with ED we employ DMRG where the total number of particles
$N$ is fixed and the resulting entanglement spectrum can be sorted according to
the corresponding numbers of particles $n$ and $N-n$ in the two partitions of
the system \cite{itensor}. This allows for the analysis of up to $L=98$ sites
at half-filling with the results shown in Fig.~\ref{fig:peakScalingOddN} where the DMRG is benchmarked against ED for $N=15$ (periodic boundary conditions).  Performing a 2-parameter fit of the DMRG data to $\tfrac{V}{t}\rvert_{\rm max} = 2 + \mathcal{A}N^{-1/\nu}$ supports a finite-size scaling form $\delta\sim L^{-1/\nu}$ $(L = 2N)$, with exponent $1/\nu \simeq 0.3$. 

\subsection{Reduction of entanglement due to particle fluctuations between subsystems}

\begin{figure}[htp]
\begin{center}
\includegraphics[width=1.0\columnwidth]{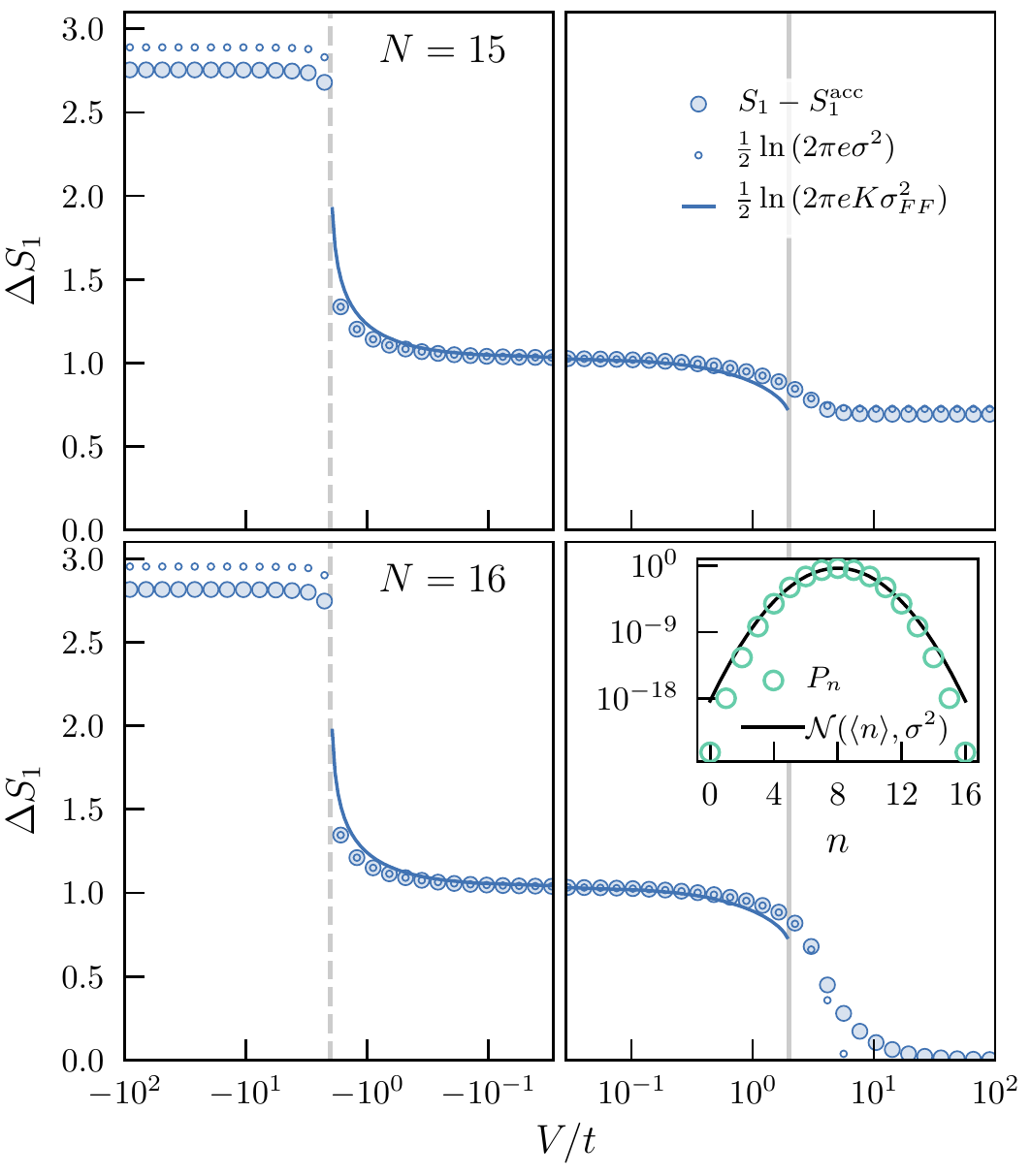}
\end{center}
\caption{Difference between the von Neumann and accessible entanglement entropies $\Delta S_1 = S_{1}-S_{1}^{\rm acc}$ (large circles) and the Shannon entropy of a Gaussian distribution, $\frac{1}{2} \ln 2 \pi e \sigma^2$ (small circles) as functions of interaction strength $V/t$. Results were determined via exact diagonalization for the ground state of Eq.~\eqref{eq:H-tV} with $N=15,16$.  The observed agreement between the large and small symbols demonstrates the rapid convergence of $P_n$ to a Gaussian distribution as seen in the inset where $\mathcal{N}(\expval{n},\sigma^2)$ is a normal distribution with the same mean and variance as $P_n$.  Solid lines are computed from the
theoretical variance of the number of fermions in region $A$ inside the
Tomonaga-Luttinger liquid phase $\sigma^2 = K \sigma^2_{FF}$, where $K$ is the Luttinger parameter computed via Eq.\eqref{eq:KtV} and $\sigma^2_{FF}$ is the exact variance for free-fermions ($V/t = 0$).}
\label{fig:deltaS1}
\end{figure}

The difference between the full and accessible von Neumann entanglement entropies, $S_1 - S_1^{\mathrm{acc}} \equiv \Delta S_1 = H_1$, is equal to the Shannon entropy $H_1 = -\sum_n P_n \ln P_n$ of the particle number distribution \cite{Klich:2008se}.  From the asymptotic results in Table~\ref{tab:Limits} we expect $\Delta S_1$ to be maximal in the limit of strong attractive interactions where it behaves like $\ln L$ as extensive particle fluctuations between spatial subsystems contribute to the entanglement.  In the opposite limit $V/t \to \infty$, we expect the difference to converge to a constant ($N$ odd) or zero ($N$ even) where repulsion strongly suppresses number fluctuations. This behavior is confirmed in Fig.~\ref{fig:deltaS1} where we show the interaction dependence of $\Delta S_1$ computed via exact diagonalization for $N=15,16$ (large circles).
%
%
Fig.~\ref{fig:deltaS1} also includes the entanglement reduction computed from the numerically determined variance of $P_n$ (small circles) under the assumption that $P_n$ is a continuous Gaussian distribution with mean $\expval{n}$ described by:
\begin{equation}
    P_n\approx\dfrac{1}{\sqrt{2\pi\sigma^2}}\exp\qty[{\frac{-(n-\langle
    n\rangle)^2}{2\sigma^2}}] \equiv \mathcal{N}\qty(\expval{n},\sigma^2)
\label{eq:PNGaussian}
\end{equation}
with associated Shannon entropy (see Eq.~\eqref{eq:gaussianEntropy}):  
\begin{equation*}
H_1 = \Delta S_1 \approx \frac{1}{2}\ln\left(2\pi e \sigma^2\right)\,.
\end{equation*}
The resulting agreement between the exact $\Delta S_1$ with the asymptotic large-$N$ result is surprisingly good over the entire range of $\lvert V/t \rvert \lesssim 2$ where $P_n$ might still be expected to retain strong signatures of discreteness at these finite values of $N$.  This is confirmed in the inset of the lower panel for $N=16$ where we compare the exact finite size probabilities $P_n$ with a Gaussian distribution $\mathcal{N}(\expval{n},\sigma^2)$ having the same mean $\langle n\rangle$ and variance $\sigma^2$ for a particular coupling $V/t = -1.5$. 

Moreover, we can quantitatively capture the interaction dependence of $\Delta
S_1$ (solid lines in Fig.~\ref{fig:deltaS1}) using the predicted Gaussian form
and variance of the number distribution at low energies within the TLL regime
(in the thermodynamic limit) using $\sigma^2=K\sigma^2_{FF}$
\cite{Klich:2008se, Song:2010eq, Song:2012cp} where the Luttinger parameter $K$
is computed using Eq.~\eqref{eq:KtV} and $\sigma^2_{FF}$ is the variance of $P_n$
for free fermions. We note that we do not include a subleading interaction
dependent term in $\sigma^2$ to prevent over-fitting.

To better understand the highly Gaussian nature of the subsystem particle number
probability distribution, we restrict to the case of even $N$, where the
symmetry $P_n$ at half-filling guarantees that $\expval{n}=N/2$ is an integer such that $\delta n=n-\langle n\rangle$ is also an integer. Using the Poisson summation formula for a Gaussian function we find:
\begin{equation}
    \sum_{\delta n=-\infty}^{\infty}e^{-\frac{(\delta n)^2}{2\sigma^2}}=\sqrt{2\pi\sigma^2}\left[1+2\sum_{\delta n=1}^{\infty}e^{-2\pi^2\sigma^2 (\delta n)^2}\right]
\label{eq:PoissonSum},
\end{equation}
where the summation on the right-hand side represents the error in the
normalization of $P_n$ which decreases with increasing variance $\sigma^2$ (the odd $N$ case is analogous~\footnote{For odd $N$, the relevant Poisson summation formula is $\sum_{\delta
n=-\infty}^{\infty}\exp\left[-(\delta
n+1/2)^2/(2\sigma^2)\right]=\sqrt{2\pi\sigma^2}\qty{1+2\sum_{\delta
n=1}^{\infty}(-1)^{\delta n} \exp\qty[-2\pi^2\sigma^2(\delta n)^2]}$.}). For the data presented in the inset of Fig.~\ref{fig:deltaS1}, the value of $\sigma^2$ is $0.772$ $(N=16)$  leading to a corresponding error of $\sim10^{-6}$. Taking the derivative of both sides of \Eqref{eq:PoissonSum} with respect to $\sigma^2$ shows that the variance of $P_n$ calculated using its expression in \Eqref{eq:PNGaussian} is well approximated by $\sigma^2$ in the same limits. 

\begin{figure}[t]
\begin{center}
\includegraphics[width=1.0\columnwidth]{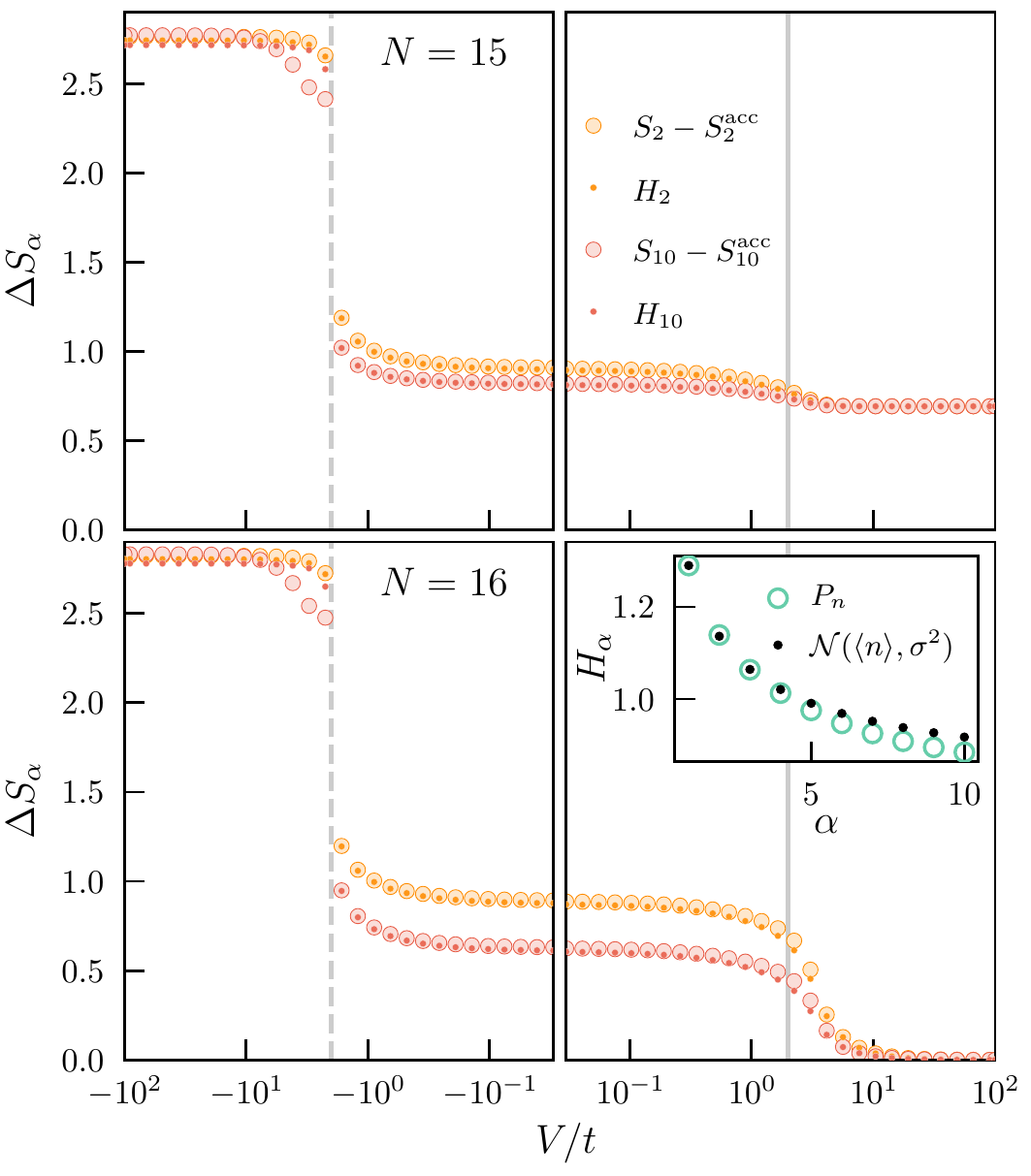}
\end{center}
\caption{Interaction dependence of the difference between the \ren and accessible entanglement entropy $\Delta S_\alpha =  S_{\alpha} - S_{\alpha}^{\mathrm{acc}}$.  Large symbols are computed via exact diagonalization for the ground state of the $t-V$ model at half-filling with a spatial partition corresponding to $L/2$ sites.  Small symbols are the classical \ren entropies $H_\alpha$ computed from $P_n$, the probability of finding $n$ particles in the spatial subregion.  We observe $H_{\alpha} \leq \Delta S_{\alpha}$ with the lower bound being nearly saturated over a wide range of interactions, but is worse for even $N$ and as $\alpha$ is increased from 2 to 10.  This is quantified in the inset which compares $H_\alpha$ computed from the exact probability distribution $P_n$ with that obtained from a Gaussian with the same mean and variance as $P_n$ for $N=16$.  
}
\label{fig:deltaS_alpha}
\end{figure}

We can extend this analysis to the case of \ren indices $\alpha > 1$ with exact digonalization results shown in Fig.~\ref{fig:deltaS_alpha}.
Here the difference between the spatial and accessible entanglement is no longer exactly equal to $H_\alpha$, the classical \ren entropy of $P_n$, but is instead given by the modified expression  $H_{1/\alpha}(\qty{P_{n,\alpha}})$ as defined in Eqs.~\eqref{eq:S_alpha_acc5} -- \eqref{eq:Pna}. However, a comparison of the large and small symbols in Fig.~\ref{fig:deltaS_alpha} indicate that $\Delta S_\alpha \approx H_\alpha$ for $\qty|V/t| \lesssim 2$. This can be understood using our observation from Fig.~\ref{fig:deltaS1} that $P_n$ is well approximated by a continuous Gaussian distribution in the TLL phase.  In this case the renormalized probability distribution $P_{n,\alpha} \approx (P_n)^\alpha/\sum_n (P_n)^\alpha$ is also Gaussian with variance $\sigma^2_{\alpha}=\sigma^2/\alpha$.  As shown in Eq.~\eqref{eq:PnaPaHa}, this has the consequence that $H_{{1}/{\alpha}}\left(\{ P_{n,\alpha}\}\right)\simeq H_\alpha$ and thus the difference $\Delta S_{\alpha} \simeq H_\alpha$.   For larger values of $\alpha$, increased deviations between $\Delta S_\alpha$ and $H_\alpha$ are observed which are quantified in the inset of of the lower panel of Fig.~\ref{fig:deltaS_alpha} that compares $H_\alpha$ computed for the exact $P_n$ with that determined from a continuous normal distribution having the same mean and variance as $P_n$ for $N=16$.  For $\alpha=10$, the effects of discreteness are amplified which can be understood by returning to Eq.~\eqref{eq:PoissonSum} with $\sigma^2_\alpha = \sigma^2/\alpha$ such that the correction term on the right hand side becomes more important as the width of the distribution is squeezed.

\begin{figure}[t]
\begin{center}
\includegraphics[width=1.0\columnwidth]{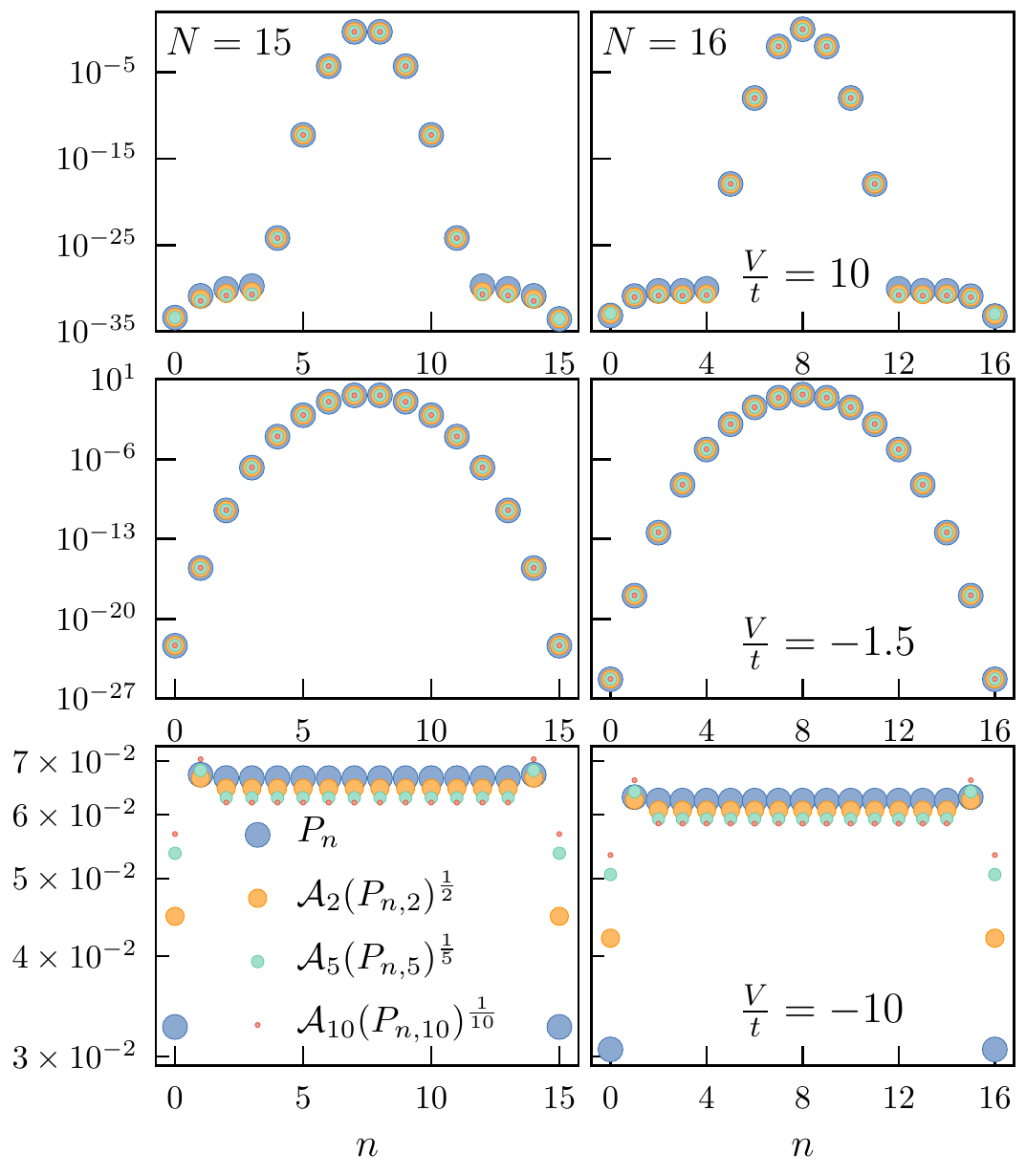}
\end{center}
\caption{Rescaling the effective probability distribution defined in Eq.~\eqref{eq:Pna} for the ground state of the $t-V$ model at half-filling demonstrates the approximate proportionality relation $P_n \sim (P_{n,\alpha})^{1/\alpha}$ for interaction strengths corresponding to the charge density wave (top row, $V/t=10$), Tomonaga-Luttinger Liquid (middle row, $V/t=-1.5$) and cluster phases (bottom row, $V/t=-10$) for \ren indices $\alpha=1,2,5,10$. $\mathcal{A}_\alpha^{-1} = \sum_n (P_{n,\alpha})^{1/\alpha}$ is a normalization constant and $P_{n,\alpha}$ is defined in Eq.~\eqref{eq:Pna}. The shape of the distributions and their connection to the physical ground states of the $t-V$ model is discussed in the text.}
\label{fig:alpha_collapse}
\end{figure}

The preceding analysis of the accessible entanglement entropy has demonstrated the importance of the specific form of the probability distribution $P_n$ and in Fig.~\ref{fig:alpha_collapse} we examine it more closely in the three phases of the $t-V$ model.  For large repulsive interactions where the system is in the CDW phase (top panel, $V/t = 10$), $P_n$ is dominated by configurations where $n = N/2$ for $N$ even and $n = (N\pm 1)/2$ for $N$ odd. In the TLL phase where $|V/t| < 2$, we have already found that $P_n$ is well described by a normal distribution (middle row, $V/t=-1.5$).  Finally, for strong attractive interactions (bottom row, $V/t=-10$), $P_n$ is nearly flat, as the ground state is a superposition of all spatial translations of the cluster of $N$ particles.  Fig.~\ref{fig:alpha_collapse} also explains the empirical observation of the semi-equality $\Delta S_{\alpha} \approx H_\alpha$ for all interaction strengths as a consequence of the proportionality $P_{n,\alpha}\sim P_n^{\alpha}$ by demonstrating the collapse of $\mathcal{A}_\alpha P_{n,\alpha}^{1/\alpha}$ to $P_n$ for different values of $\alpha$, where $\mathcal{A}_\alpha$ is a normalization factor.

\begin{figure}[t]
\begin{center}
\includegraphics[width=1.0\columnwidth]{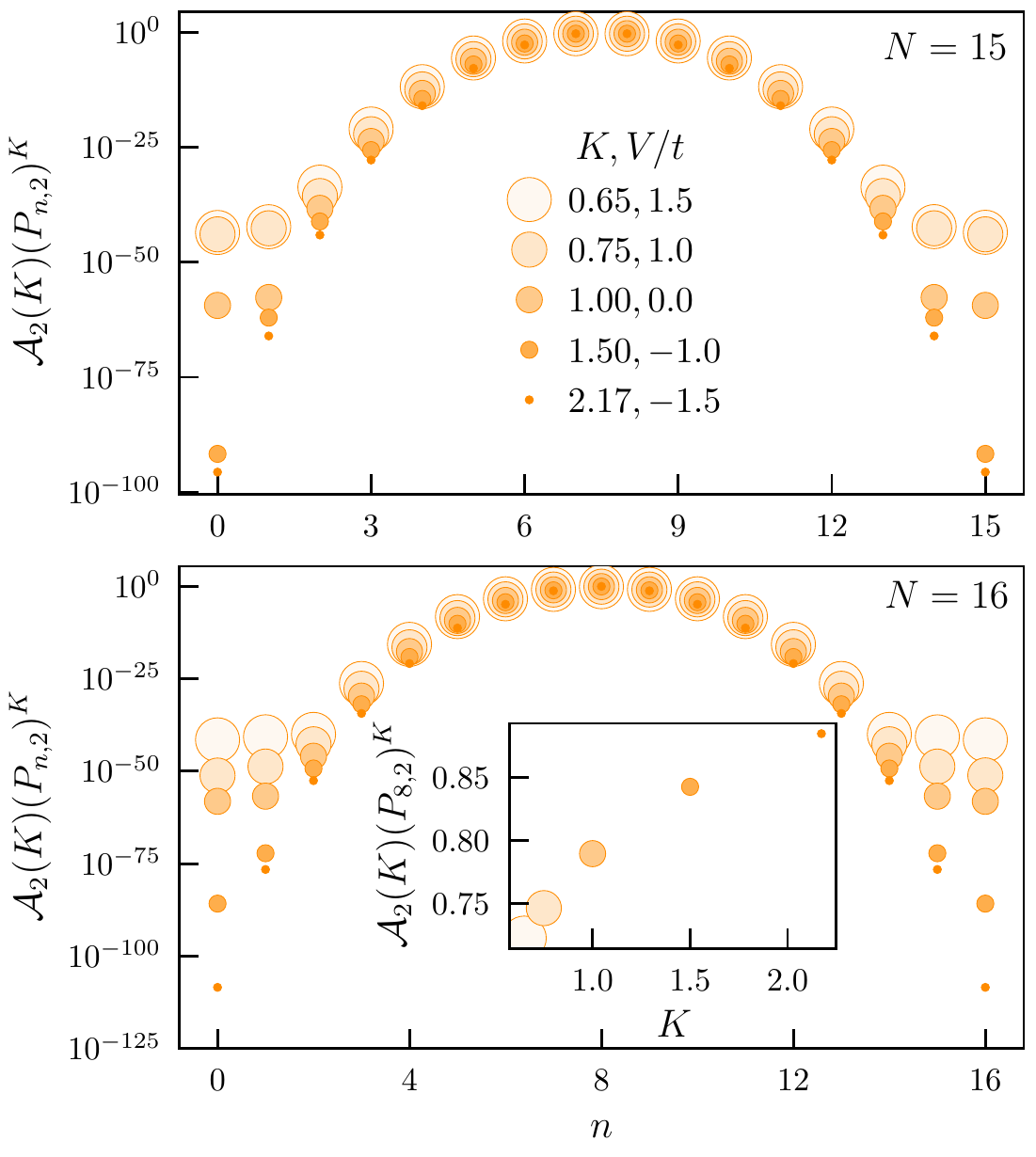}
\end{center}
\caption{The effective probability distribution $P_{n,\alpha=2}$ for the ground state of the $t-V$ model at half-filling and for different interaction strengths $V/t$ in the TLL phase is  rescaled as $[P_{n,2}]^{K}$, were $K$ is the corresponding Luttinger parameter computed from Eq.~\eqref{eq:KtV}.  While the probabilities seem to show collapse near the middle of the
distribution where $n \simeq N/2$, the inset shows strong additional $K$ dependence of the
probability for fixed particle number $n=8$ in $A$. As discussed in the text,
this lack of collapse is due to the subleading interaction dependent
corrections to the asymptotic scaling of the variance $\sigma^2$ in
Eq.~\eqref{eq:sigma2prediction}}
\label{fig:K_collapse}
\end{figure}

Motivated by the observation of the $\alpha$-collapse of the effective probability distribution $P_{n,\alpha}$, we test another asymptotic result: in the TLL phase the variance $\sigma^2_\alpha$ is expected to approach the value of $K\sigma^2_{FF}/\alpha$. This means that for a fixed $\alpha$, we expect the asymptotically Gaussian distribution $P_{n,\alpha}(K)$ for a given interaction strength $K$ to be proportional to $[P_{n,\alpha}(K=1)]^{1/K}$. This prediction is investigated in Fig.~\ref{fig:K_collapse}, where we set $\alpha=2$ and consider different interaction strengths in the TLL phase where we have used Eq.~\eqref{eq:KtV} to convert between $V/t$ and the Luttinger parameter $K$. 
%
%
On a semi-logarithmic scale, the results suggests data collapse to $P_{n,2}(K=1)$ near the middle of the distributions corresponding to $n \simeq N/2$ particles in the subregion. However, a linear-linear analysis at $n=N/2$ indicates deviations, as illustrated in the inset of  Fig.~\ref{fig:K_collapse} (bottom panel). This can be understood by considering higher order corrections to the asymptotic dependence of $\sigma^2$ on $K$, \emph{e.g.}, for $\alpha=1$, the variance of $P_n$ is given by~\cite{Song:2010eq} 
\begin{equation}
    \sigma^2 \simeq \frac{K}{\pi^2}\ln\ell  + a_1 - \frac{a_2(-1)^\ell}{\ell^{2K}}\ ,   
\label{eq:sigma2prediction}
\end{equation}
where $a_1$ and $a_2$ are $K$-dependent constants and $\ell$ is the macroscopic size of the spatial subregion. We have tested that a more faithful rescaling of the distributions with $\sigma^2/\sigma^2_{FF}$ instead of $K$ leads to improved data collapse, especially if $\sigma^2$ and $\sigma^2_{FF}$ are calculated by fitting the middle portion of the distribution to a Gaussian function, instead of requiring them to be the variance of the corresponding distribution.

Until now, we have focused on the case of a half-filled lattice: $ N = L/2$.  A more general result for the scaling of the variance of the particle number distribution $\sigma^2$ (fluctuation entanglement) in the TLL phase for a system of size $L,N \gg 1$ but with a finite filling fraction $N/L$ is given by \cite{Song:2012cp,Calabrese:2011fj}
%
%
%
\begin{equation}
\sigma^2\approx\mathcal{F}(N, \ell)=\frac{K}{2\pi^2}\ln\left[\left(\frac{\pi N\ell} {L}\right)^2+1\right]
\label{eq:FuctuationeEtanglement}.
\end{equation}
In order to compute $K$ above we note that
away from half-filing,  Eq.~\eqref{eq:KtV} is no longer valid and the $V/t$
dependence of the Luttinger parameter must be determined via a full
numerical solution of the Bethe ansatz equations for the
corresponding XXZ model at each filling fraction $N/L$.  For $\ell \gg 1$, the above expression simplifies to the known asymptotic result $\sigma^2\approx({K}/{\pi^2})\ln\qty(k_{\rm{F}} \ell a_0)$\cite{Klich:2008se}, where $k_{\rm{F}}={\pi N}/({L a_0})$ is the Fermi momentum and $a_0$ is a microscopic length scale. 

\begin{figure}[t]
\begin{center}
\includegraphics[width=1.0\columnwidth]{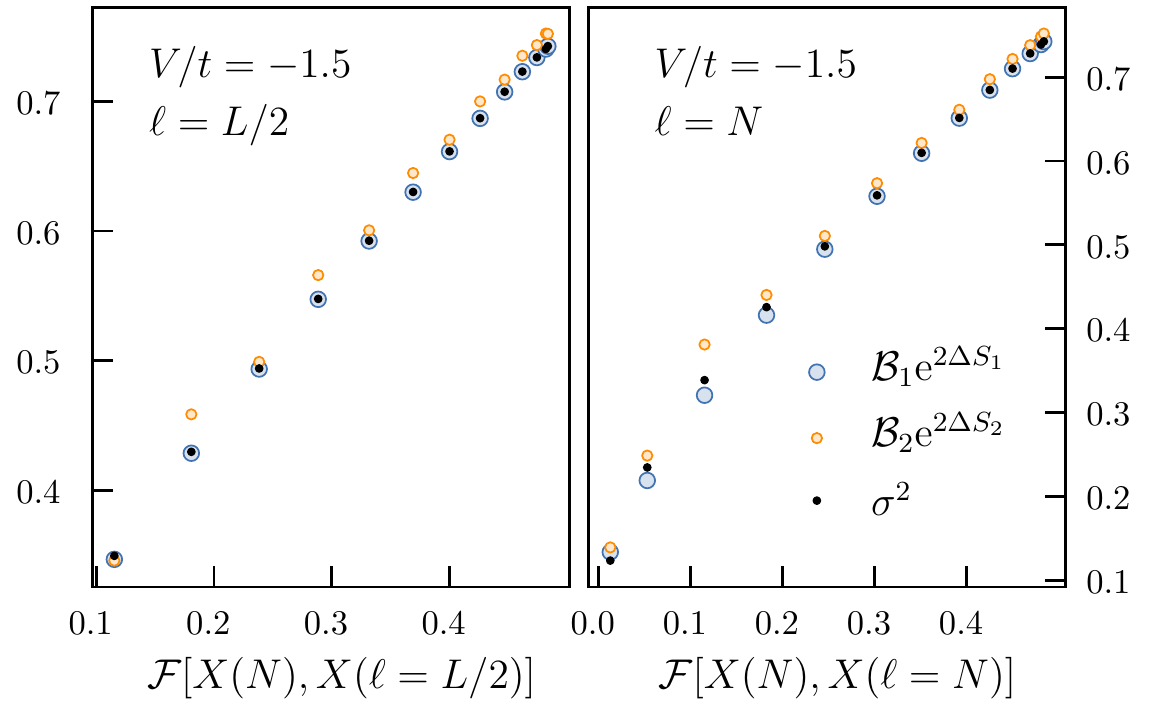}
\end{center}
\caption{The filling fraction dependence of the variance $\sigma^2$  (black
    dots) of the  particle number distribution $P_n$ compared to that computed
    from the difference between the (exponentiated) accessible and spatial
    entanglement entropies for $\alpha=1$ (large circles) and $\alpha=2$
    (small circles) using Eq.~\eqref{eq:sigma2DeltaS}.  Exact diagonalization
    results in the ground state of the $t-V$ model with $V/t = -1.5$ and $L=28$
    demonstrate consistency with the predicted asymptotic scaling function
    $\mathcal{F}(N, \ell)$ given in Eq.~\eqref{eq:sigma2prediction} where we have fixed $\ell = L/2$ (left panel) or set $\ell = N$ (right panel) and changed $N$ from $2\dots14$.  In order to reduce finite size effects, both of the partition size $\ell$ and the total number of particles $N$  are replaced by their corresponding chord length $X(\ell)=(L/\pi)\sin(\pi\ell/L)$ and $X(N)$, respectively.  The constants $\mathcal{B}_1=\left(\pi e\right)^{-1}$ and $\mathcal{B}_2=\left(2\pi \right)^{-1}$ are used to rescale the entanglement reduction to obtain a prediction for the variance $\sigma^2$.
}
\label{fig:fillingFractionDependence}
\end{figure}

In Fig.~\ref{fig:fillingFractionDependence}, we explore the scaling prediction of  Eq.~\eqref{eq:FuctuationeEtanglement}.  
%
%
In the left panel we increase the number of fermions $N$ in a system with a fixed system size $L = 28$ and partition size $\ell=L/2 = 14$, while in the right panel we set $L = 28$ but grow $N$ and $\ell$ together, \emph{i.e.}\ $\ell=N$.  To take into account the finite size and periodic boundary conditions in our exact diagonalization calculations we replace $\ell$ with the chord length $X(\ell)=(L/\pi)\sin(\pi\ell/L)$ and similarly $N$ with $X(N)$. We observe that the numerical results are consistent with with Eq.~\eqref{eq:FuctuationeEtanglement} for a modest system size. 

We also investigate the prediction that $P_n$ should remain a Gaussian distribution, even away from half-filling by solving for $\sigma^2$ using Eq.~\eqref{eq:gaussianEntropy} we find 
\begin{equation}
\sigma^2=\mathcal{B_{\alpha}}\exp(2\Delta S_{\alpha})
\label{eq:sigma2DeltaS},
\end{equation}
%
%
where $\mathcal{B}_{\alpha} \equiv \alpha^{\left(1-\alpha\right)^{-1}}/{\pi}$ and taking the appropriate limit yields $\mathcal{B}_1=\left(\pi e\right)^{-1}$. For $\alpha =1$,  we expect that Eq.~\eqref{eq:sigma2DeltaS} should asymptotically hold, as long as $P_n$ is a Gaussian distribution with variance $\sigma^2$. This is confirmed by the agreement between the large circles and filled black dots in Fig.~\ref{fig:fillingFractionDependence}. 

For $\alpha\ne1$, the validity of Eq.~\eqref{eq:sigma2DeltaS} requires that both $P_{n,\alpha}$ is Gaussian, and that its variance $\sigma^2_{\alpha}$ behaves as $\sigma^2_{\alpha}=\sigma^2/\alpha$. In Fig.~\ref{fig:fillingFractionDependence} we see that this is almost the case for $\alpha = 2$, where the small deviations can be attributed to the squeezed variance of $\sigma^2_2=\sigma^2/2$ compared to $\sigma^2$ as discussed above.  In other words, even if the most relevant part of $P_n$ takes the form of a discrete Gaussian distribution, the value of the parameter $\sigma^2$ in the exponent of $\exp\qty[-{(\delta n)^2}/{(2\sigma^2)}]$ can only approximately represent the variance of the true distribution, with an accuracy that increases with $\sigma^2$ as derived from Eq.~\eqref{eq:PoissonSum}. The weak oscillations which appear in Fig.~\ref{fig:fillingFractionDependence} for $\alpha=2$ are due to the same anomalous scaling corrections that appear in the \ren entanglement entropy with $\alpha>1$\cite{Calabrese:2010cm}.

\section{Discussion}

In this paper we have presented a systematic study of how nearest-neighbor interactions affect the amount of operationally accessible entanglement that could be extracted from the ground state of a system of one-dimensional spinless lattice fermions where the total number of particles is fixed.  The existence of this superselection rule (fixed $N$) limits the set of physical operations that can be performed with the result that the entanglement entropy under a spatial mode bipartition provides an absolute upper bound on the accessible entanglement. We have derived analytic results for the von Neumann ($\alpha = 1$) and generalized \ren ($\alpha \ne 1$) accessible entanglement in a few special cases (see Table~\ref{tab:Limits}).  In the limit of strong attractive interactions, the ground state is a superposition of all translations of a single cluster of $N$ fermions and the accessible entanglement is reduced by $\ln N$ from the spatial entanglement saturating at a constant for large $N$. For strong repulsive interactions at half filling, the ground state is a superposition of possible density waves commensurate with the number of sites and the accessible entanglement is equal to the spatial entanglement for even $N$ (no reduction), while it is reduced by a constant term to zero for odd $N$. Finally, exactly at the first order phase transition at $V/t = -2$, the ground state is an equal weight superposition of all possible fermion occupation states and the accessible entanglement is identically zero for all filling fractions and system sizes. This constitutes the maximal possible reduction, with all of the spatial entanglement entropy, which scales as the logarithm of the subsystem size, being due to particle fluctuations. This result highlights the importance of understanding the role of classical number fluctuations in itinerant many-body systems when using entanglement entropy as a phase diagnostic. The drastic reduction in entanglement after projection into fixed particle number subsectors is reminiscent of Yang's $\eta$-paired state \cite{Yang:1989eu} under the quantum disentangled liquid diagnostic \cite{Grover:2011km, Garrison:2017mn, Veness:2017uu} which involves a partial projection onto spin degrees of freedom.

Within the Tomonaga-Luttinger liquid phase $\qty|V/t| < 2$ the asymptotic form
of the particle number distribution $P_n$ is known to be Gaussian with a
variance that scales as $\sigma^2 \simeq  (K/\pi^2) \ln \ell$ for $\ell \gg 1$
\cite{Klich:2008se, Song:2010eq}. $\sigma^2$ is parametrically large enough
within the quantum liquid (especially for attractive interactions) that the
discreteness of the underlying $P_n$ distribution can be neglected.
Fluctuations in this regime are not the only factor controlling
entanglement, and the presence of
interactions ensures that the spatial entanglement entropy is reduced by the superselection
rule only by a subleading double logarithm. Thus the fermionic Luttinger liquid
at half-filling can be considered a useful entanglement resource.

At the continuous quantum phase transition between the TLL and charge density wave,  we observe a global maxima in the accessible entanglement which demonstrates a susceptibility-like scaling consistent
with the known thermodynamic limit critical value of $V/t = 2$. Confirmation of
this scaling, especially away from half-filling, would require studying larger
system sizes than considered here. Ultimately we are limited by the well known
difficulties of DMRG when investigating ground states with a large amount of entanglement (here scaling like $\sim \ln(L/2)$ for $\ell = L/2$) near the critical point, especially with periodic boundary conditions as considered here.  There are many natural extensions utilizing DMRG with access to quantum numbers describing subregion particle occupation numbers, including investigating the effects of boundary conditions, different partition sizes, and extended range interactions.

The difference between the von Neuman ($\alpha=1$) accessible and spatial entanglement entropies, $\Delta S_1$, is exactly given by the Shannon entropy $H_1$ of the corresponding particle number distribution $P_n$ \cite{Klich:2008se}. A direct \ren generalization of this relation to $\alpha \neq 1$ is not true \cite{Barghathi:2018oe}, \emph{i.e.},\ $\Delta S_{\alpha} \neq H_{\alpha}$.  However, a sufficient condition for such a generalization is that $P_{n,\alpha} \propto (P_n)^\alpha$ where the constant of proportionality can be dependent on $\alpha$ but not on $n$.  This is equivalent to requiring that the trace of the projected reduced density matrix raised to the power $\alpha$, $\Tr\, \rho^\alpha_{A_n}$, be independent of $n$. This is always the case asymptotically for $\ell \gg 1$ when the number fluctuations are Gaussian  with a variance $\sigma^2_{\alpha}$ that is inversely proportional to $\alpha$, $(|V/t| < 2)$, but we find it to be approximately satisfied throughout the phase diagram, 
even away from half-filling.  However, deviations occur in the limit of strong attractive interactions, or when $\alpha \gg 1$. In this case, large $\alpha$ always tends to reduce the variance of the effective distribution $P_{n,\alpha}$ and thus for finite size systems, the discreteness of the physical number of particles in spatial subregion $A$ can further spoil the semi-equality between $\Delta S_\alpha$ and $H_\alpha$.  The fact that $\Delta S_\alpha \approx H_\alpha$ when $V/t \gg 1$ is a consequence of the separation of scales in this limit where $P_n$ is dominated by configurations with $n \simeq N/2$.

This result accentuates the importance of the superselection rule in reducing accessible entanglement and provides a direct route towards the experimental measurement of $\Delta S_\alpha$ in systems of ultracold atoms via a quantum gas microscope \cite{Bakr:2009ct}.

Many open questions remain, and having demonstrated the utility of the operationally accessible entanglement in an exactly solvable model, it is natural to ask what this quantity can tell us about non-integrable models in one dimension as well interacting fermions and soft-core bosons in higher dimensions.  In the latter case, the support of $P_n$ is no longer bounded by the number of sites in the spatial subregion, and the study of large systems could be performed via quantum Monte Carlo \cite{Herdman:2014ey} simulations. Recent work validating the connection between subregion particle fluctuations and spatial entanglement in a non-equilibrium setting \cite{Gruber:2019kt} could also be extended to probe how superselection rules may affect the dynamics of accessible entanglement after a quantum quench.   

From a quantum information perspective, it seems important to further explore how the accessible entanglement relates to the plethora of measures \cite{Barnum:elViola:2005,Schwaiger:2015id,Sauerwein:2015ka,LoFranco:2018dd,Benatti:2014} which do not directly include physical restrictions on $N$, but aim to quantify the technologically useful quantum correlations encoded in interacting and indistinguishable itinerant quantum particles.


 \section{Acknowledgments}

We benefited from discussion with C.~M.~Herdman, F.~Heidrich-Meisner, I.~Klich and  M. Stoudenmire.  This research was supported in part by the National Science Foundation (NSF) under award No.~DMR-1553991 (A.D.). All computations were performed on the Vermont Advanced Computing Core supported in part by NSF award No.~OAC-1827314.


\appendix
\section{Ground state of the $t-V$ model for $V/t = -2$}
\label{Appendix:A}
Consider the Hamiltonian of the $t-V$ model given in Eq.~\eqref{eq:H-tV} at the special interaction strength $V=-2t$ corresponding to the first order phase transition:
\begin{equation}
    H = -t \sum_{i=1}^L (c_{i}^\dagger c_{i+1}^{\phantom{\dagger}} + c_{i+1}^\dagger c_{i}^{\phantom{\dagger}}) -2t \sum_{i=1}^L n_i n_{i+1}
\end{equation}
where we assume periodic boundary conditions for $N$ even and antiperiodic
boundary conditions for $N$ odd.  

\subsection{Fermion occupation basis}

We study the effect of $H$ in the $N$ fermion occupation basis $\{\ket{\psi_a}\}$, where the index $a$ runs over all of the $L\choose N$ possible configurations.  For example, for $N=2$ and $L=4$ there are six such states: $\ket{\psi_a} \in \{\ket{1100}, \ket{1010}, \ket{1001}, \ket{0110}, \ket{0101}, \ket{0011}\}$. 

Starting with the potential operator $\mathcal{V} \equiv -2t \sum_{i=1}^L n_i n_{i+1}$ which is diagonal in this basis, we have
\begin{equation}
    \mathcal{V}\ket{\psi_a} =-2t\, n^{(11)}_a\ket{\psi_a}\, ,
    \label{eq:Vpsia}
\end{equation}
where $ n^{(11)}_a$ counts the number of bonds connecting two occupied sites in the state $\ket{\psi_a}$. The hopping operator $\mathcal{T} \equiv -t \sum_{i=1}^L (c_{i}^\dagger c_{i+1}^{\phantom{\dagger}} + c_{i+1}^\dagger c_{i}^{\phantom{\dagger}})$ turns $\ket{\psi_a}$ into a superposition of all the states $\ket{\psi_b}$ connected to $\ket{\psi_a}$ by moving one particle to a neighboring empty site. We can write: 
\begin{equation}
    \mathcal{T}\ket{\psi_a} =-t \sum_{b \in \mathsf{S}_a}\ket{\psi_b}\, ,
    \label{eq:Tpsia}
\end{equation}
where $\mathsf{S}_a$ is the resulting index set of occupation states $\ket{\psi_b}$,
i.e. $b \in \mathsf{S}_a \iff \matrixel{\psi_b}{\mathcal{T}}{\psi_a} \ne 0$.  
The cardinality of $\mathsf{S}_a$ is
\begin{align}
    \mathrm{card}(\mathsf{S}_a) &\equiv \sum_{b \in \mathsf{S}_a} 1\nonumber \\
    &= n^{(10)}_a+n^{(01)}_a \nonumber \\
    &= 2N - 2n_a^{(11)}, 
\label{eq:cardSa}
\end{align}
where $n^{(10)}_a$ ($n^{(01)}_a$) counts the number of occupied-empty
(empty-occupied) bonds in $\ket{\psi_a}$ and in the last line we have used the fact
that the total number of particles on a ring is (independent of the index $a$)
\begin{equation}
 N=n^{(11)}_a+(n^{(10)}_a+n^{(01)}_a)/2\,.
\label{eq:Nring}
\end{equation}
A general matrix element in the fermion occupation basis is given by:
\begin{equation}
    \matrixel{\psi_c}{\mathcal{T}}{\psi_a} = -t
    \begin{cases}
        1 & c \in \mathsf{S}_a \\
        0 & \text{otherwise}
    \end{cases}
\label{eq:Tmatrixelemnts}
\end{equation}
which is guaranteed to be real, thus 
\begin{equation}
\matrixel{\psi_c}{\mathcal{T}}{\psi_a} = 
\matrixel{\psi_a}{\mathcal{T}}{\psi_c} \Rightarrow c \in \mathsf{S}_a \iff a \in
    \mathsf{S}_c.
\label{eq:setSwitch}
\end{equation}
This is a useful result that can be used to swap the order of restricted and
un-restricted summations.

Let us know consider the action of $\mathcal{T}$ 
    on a general state $\ket{\Psi} = \sum_a \mathcal{C}_a \ket{\psi_a}$ where $\mathcal{C}_a \in \mathds{C}$:
\begin{align}
    \mathcal{T}\ket{\Psi} &= -t \sum_a \mathcal{C}_a \sum_{b \in \mathsf{S}_a} \ket{\psi_b} \nonumber \\
                          &= -t \sum_c \ket{\psi_c} \sum_a \mathcal{C}_a
                          \sum_{b \in \mathsf{S}_a} \braket{\psi_c}{\psi_b} \nonumber  \\ 
                          &= -t \sum_c \ket{\psi_c}\left[ \sum_a \mathcal{C}_a
                          \sum_{b \in \mathsf{S}_a} \delta_{c,b}\right] 
    \label{eq:TPsi1}
\end{align}
where we have inserted a resolution of the identity operator $\sum_c
\ket{\psi_c}\bra{\psi_c} = \mathds{1}$ into the second line. Now, $\sum_{b \in
\mathsf{S}_a}\delta_{c,b} \ne 0 \iff c \in \mathsf{S}_a$ and using
\Eqref{eq:setSwitch} we can write
\begin{equation}
    \sum_a \mathcal{C}_a \sum_{b \in \mathsf{S}_a} \delta_{c,b} = \sum_{a \in S_c}
    \mathcal{C}_a \,.
\end{equation}
Substituting into \Eqref{eq:TPsi1} above and relabelling $a \leftrightarrow c$
leads to the general result:
\begin{equation}
    \mathcal{T}\ket{\Psi} = -t \sum_a \sum_{c \in \mathsf{S}_a} \mathcal{C}_c  \ket{\psi_a}.
    \label{eq:TPsi2}
\end{equation}
Written in this form, we can combine \Eqref{eq:TPsi2} with
Eqs.~(\ref{eq:Vpsia}) and (\ref{eq:cardSa}) to compute the action of the full Hamiltonian at $V=-2t$ on $\ket{\Psi}$:
\begin{align}
    H \ket{\Psi} &= -t \sum_a \left[\sum_{c \in \mathsf{S}_a} \mathcal{C}_c + 2
    n^{(11)}_a \mathcal{C}_a \right] \ket{\psi_a} \nonumber \\ 
                 &= -2t N \ket{\Psi} - t \sum_a \sum_{c \in \mathsf{S}_a}
                 \qty(\mathcal{C}_c - \mathcal{C}_a)\ket{\psi_a}\,.
\label{eq:HPsi}
\end{align}

\subsection{The Flat State}

From \Eqref{eq:HPsi} it is immediately apparent that the flat state
\begin{equation}
\ket{\Psi_0} = \frac{1}{\sqrt{{L}\choose{N}}} \sum_a \ket{\psi_a}
    \label{eq:Psiflat}
\end{equation}
is an eigenstate of $H$ with energy $-2 t N$. To prove that $\ket{\Psi_0}$ is
indeed the ground state, we consider matrix elements of the shifted operator $H' = H +
2 t N$ for a general state $\ket{\Psi}$ expanded in the fermion occupation
basis:
\begin{align}
    \expval{H'}{\Psi} &= -t \sum_{a,b} \sum_{c \in \mathsf{S}_a}
    \qty(\mathcal{C}_c - \mathcal{C}_a) \braket{\psi_b}{\psi_a}
    \mathcal{C}^\ast_b \nonumber \\
                      &= t \sum_a \sum_{c \in \mathsf{S}_a}\qty(
                      \qty|\mathcal{C}_a|^2 - \mathcal{C}_a^\ast \mathcal{C}_c)
                      \nonumber \\
                      &= t \sum_a \sum_{c \in \mathsf{S}_a}\qty(
                      \qty|\mathcal{C}_c|^2 - \mathcal{C}_c^\ast \mathcal{C}_a)
\label{eq:HpExpValue}
\end{align}
where we have swapped the summations (and relabelled) in the last line making
use of \Eqref{eq:setSwitch}. Now, we can rewrite the matrix element as:
\begin{align}
    \expval{H'}{\Psi} &= \frac{t}{2} \sum_a \sum_{c \in \mathsf{S}_a} \qty(
                      \qty|\mathcal{C}_a|^2 - \mathcal{C}_a^\ast \mathcal{C}_c
+ \qty|\mathcal{C}_c|^2 - \mathcal{C}_c^\ast \mathcal{C}_a) \nonumber \\
                      &= \frac{t}{2} \sum_a \sum_{c \in \mathsf{S}_a}
                      \qty|\mathcal{C}_a - \mathcal{C}_c|^2 \ge 0. 
\label{eq:HpExpValue2}
\end{align}
Thus $H'$ is a positive operator and the flat state $\ket{\Psi_0}$ is the
ground state of $H$ at $V = -2t$ for fixed $N$. 

\FloatBarrier

\bibliography{refs}

\end{document}